\documentclass[sigconf]{acmart}

\usepackage{multirow}
\usepackage{appendix}
\usepackage{subfig}
\usepackage{graphicx}
\usepackage{grffile}

\newcounter{subcopyrightbox@save}
\usepackage[font=bf]{caption}
\usepackage{color, url}
\usepackage{xspace} 
\usepackage{mathrsfs}
\usepackage{thmtools, thm-restate}
\usepackage{hyperref}
\usepackage{epstopdf}
\usepackage{balance}
\usepackage{bm}
\usepackage{rotating}
\usepackage{enumitem}
\setlist{nolistsep}

\usepackage{algorithm}
\usepackage{algorithmic}

\newcommand{\argmax}{\operatornamewithlimits{argmax}}
\newcommand{\argmin}{\operatornamewithlimits{argmin}}

\newcommand{\myparatight}[1]{\smallskip\noindent{\bf {#1}:}~}

\graphicspath{ {../fig/} }

\begin{document}

\copyrightyear{2017} 
\acmYear{2017} 
\setcopyright{acmcopyright}
\acmConference{ACSAC 2017}\acmBooktitle{2017 Annual Computer Security Applications Conference, December 4--8, 2017, San Juan, PR, USA}\acmPrice{15.00}\acmDOI{10.1145/3134600.3134606}
\acmISBN{978-1-4503-5345-8/17/12}


\title{Mitigating Evasion Attacks to Deep Neural Networks via Region-based Classification}

\author{Xiaoyu Cao}
\orcid{0000-0002-8731-2706}
\affiliation{Duke University}
\email{xiaoyu.cao@duke.edu}
\author{Neil Zhenqiang Gong}
\affiliation{Duke University}
\email{neil.gong@duke.edu}


\begin{abstract}
Deep neural networks (DNNs) have transformed several artificial intelligence research areas including computer vision, speech recognition, and natural language processing. However, recent studies demonstrated that DNNs are vulnerable to adversarial manipulations at testing time. Specifically, suppose we have a testing example, whose label can be correctly predicted by a DNN classifier. An attacker can add a small carefully crafted noise to the testing example such that the DNN classifier predicts an incorrect label, where the crafted testing example is called \emph{adversarial example}. Such attacks are called \emph{evasion attacks}. Evasion attacks are one of the biggest challenges for deploying DNNs in safety and security critical applications such as self-driving cars.  
In this work, we develop new methods to defend against evasion attacks. Our key observation is that adversarial examples are close to the classification boundary. Therefore, we propose \emph{region-based classification} to be robust to adversarial examples. For a benign/adversarial testing example, we ensemble information in a hypercube centered at the example to predict its label. Specifically, we sample some data points from the hypercube centered at the testing example in the input space; we use an existing DNN to predict the label for each sampled data point; and we take a majority vote among the labels of the sampled data points as the label for the testing example. 
 In contrast, traditional classifiers are \emph{point-based classification}, i.e., given a testing example, the classifier predicts its label based on the testing example alone. Our evaluation results on MNIST and CIFAR-10 datasets demonstrate that our region-based classification can significantly mitigate evasion attacks without sacrificing classification accuracy on benign examples. 
Specifically, our region-based classification achieves the same classification accuracy on testing benign examples as point-based classification, but our region-based classification is significantly more robust than point-based classification to various evasion attacks. 
\end{abstract}

\maketitle

\section{Introduction}

\begin{figure}[!t]
\centering
\includegraphics[width=0.45 \textwidth]{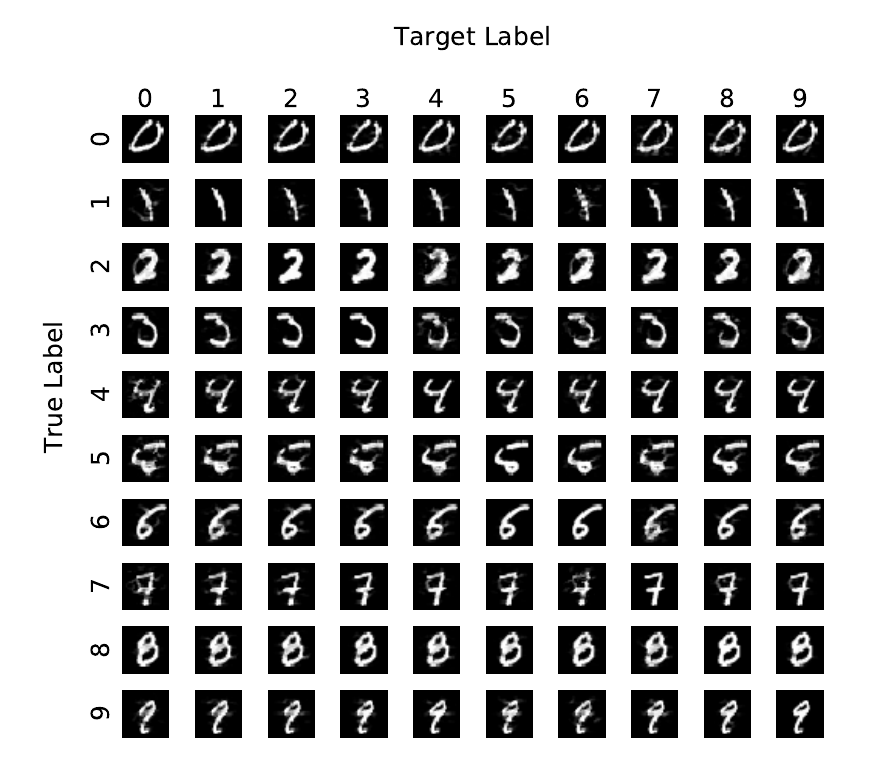}
\caption{Adversarial examples generated by an evasion attack proposed by Carlini and Wagner~\cite{CarliniSP17}.}
\label{example}
\end{figure}

Deep neural networks (DNNs) are unprecedentedly effective
at solving many challenging artificial intelligence problems such as image recognition~\cite{Krizhevsky12ImageNet}, speech recognition~\cite{hinton2012deep}, natural language processing~\cite{mikolov2013efficient}, and playing games~\cite{silver2016mastering}. For instance, DNNs can recognize images with accuracies that are comparable to human~\cite{Krizhevsky12ImageNet}; and they can outperform the best human Go players~\cite{silver2016mastering}.

However, researchers in various communities--such as security, machine learning, and computer vision--have demonstrated that DNNs are vulnerable to attacks at testing time~\cite{szegedy2013intriguing,goodfellow2014explaining,Papernot16,deepfool,liu2017delving,CarliniSP17,PracticalBlackBox17}. For instance, in image recognition, an attacker can add a small noise to a {testing example} such that the example is misclassified by a DNN classifier. The  testing example with noise is called \emph{adversarial example}~\cite{szegedy2013intriguing}. In contrast, the original example is called \emph{benign example}. Usually, the noise is so small such that, to human, the benign example and adversarial example still have the same label. Figure~\ref{example} shows some adversarial examples for digit recognition in the MNIST dataset. The adversarial examples were generated by the state-of-the-art evasion attacks proposed by Carlini and Wagner~\cite{CarliniSP17}. We use the same DNN classifier as the one used by them. The examples in the $i$th row have true label $i$, while the examples in the $j$th column are predicted to have label $j$ by the DNN classifier, where $i,j=0, 1, \cdots, 9$.

Evasion attacks limit the use of DNNs in safety and security critical applications such as self-driving cars. The adversarial examples can make self-driving cars make unwanted decisions. For instance, one basic capability of self-driving cars is to automatically recognize stop signs and traffic lights. Suppose an adversary creates an adversarial stop sign, i.e., the adversary adds several human-imperceptible dots to a stop sign,  such that the self-driving car does not recognize it as a stop sign. As a result, self-driving cars will not stop at the stop sign and may collide with other cars, resulting in severe traffic accidents.

To defend against evasion attacks, Goodfellow et al.~\cite{goodfellow2014explaining} proposed to train a DNN via augmenting the training dataset with adversarial examples, which is known as \emph{adversarial training}. Specifically, for each training benign example, the learner generates a training adversarial example using evasion attacks. Then, the learner uses a standard algorithm (e.g., back propagation) to learn a DNN using the original training benign examples and the corresponding adversarial examples. Adversarial training is not robust to adversarial examples that are unseen during training.
Papernot et al.~\cite{Papernot16Distillation} proposed a distillation based method to train DNNs. 
Carlini and Wagner~\cite{CarliniSP17} demonstrated that their attacks can still achieve 100\% success rates for DNNs trained with distillation. 
Carlini and Wagner~\cite{CarliniSP17} concluded that all defenses should be evaluated against state-of-the-art evasion attacks, i.e., the attacks proposed by them at the time of writing this paper. For simplicity, we call their attacks CW.

\myparatight{Our work} We propose a new defense method called \emph{region-based classification}. Our method can reduce success rates and/or increase the noise added into adversarial examples for various evasion attacks. First, we performed a measurement study about adversarial examples. We trained a 10-class DNN classifier on the standard MNIST dataset to recognize digits in images. The DNN has the same architecture as the one used by Carlini and Wagner~\cite{CarliniSP17}. Suppose we have a testing digit 0. We use a CW attack to generate an adversarial example for each target label 1, 2, $\cdots$, 9. Each example is represented as a data point in a high-dimensional space. For each adversarial example, we sample 10,000 data points from a small \emph{hypercube} centered at the adversarial example in the high-dimensional space. We use the DNN classifier to predict labels for the 10,000 data points. We found that a majority of the 10,000 data points are still predicted to have label 0. Our measurement results indicate that 1) the adversarial examples are close to the classification boundary, and 2) ensembling information in the hypercube around an adversarial example could correctly predict its label. 

Second, based on our measurement results, we propose a \emph{region-based classification}. In our region-based classification, we learn a DNN classifier using standard training algorithms. When predicting label for a testing example (benign or adversarial), we sample $m$ data points uniformly at random from the hypercube that is centered at the testing example and has a length of $r$. We use the DNN classifier to predict label for each sampled data point. Finally, we predict the label of the testing example as the one that appears the most frequently in the sampled data points, i.e., majority vote among the sampled data points. To distinguish our region-based classification with traditional DNN classification, we call traditional DNN \emph{point-based classification}.

One challenge for our region-based classification is how to determine the length $r$ of the hypercube. $r$ is a critical parameter that controls the tradeoff between robustness to adversarial examples and classification accuracy on benign examples. To address the challenge, we propose to learn the length $r$ using a validation dataset consisting of only benign examples. We do not use adversarial examples because the adversarial examples used by the attacker may not be accessible to the defender.
Our key idea is to select the maximal length $r$ such that the classification accuracy of our region-based classification on the validation dataset is no smaller than that of the standard point-based DNN classifier. We propose to select the maximal possible length,  so an adversarial example needs a larger noise to move further away from the classification boundary in order to evade our region-based classification. 

Third, we evaluate our region-based classification using two standard image recognition datasets, MNIST and CIFAR-10. We evaluate our region-based classification against six targeted evasion attacks and seven untargeted evasion attacks. First, our evaluation results demonstrate that our region-based classification achieves the same classification accuracy on testing benign examples with the standard point-based classification. However, adversarial training and distillation sacrifice classification accuracy.
Second, our region-based classification is significantly more robust than existing methods. For instance,  
the targeted CW attacks have less than 20\% and 7\% success rates on the MNIST and CIFAR-10 datasets, respectively. In contrast, for standard point-based classification, adversarial training, and defensive distillation, the targeted CW attacks achieve 100\% success rates on both datasets. Third, we consider an attacker strategically adapts an existing attack to our region-based classification. In particular, the attacker adds more noise to an adversarial example generated by an attack to move it further away from the classification boundary. Our results demonstrate that our region-based classification can also effectively defend against such adapted attacks. In particular, the largest success rate that the adapted attacks can achieve on the MNIST dataset is 64\%, when the attacker doubles the noise added to adversarial examples. 

We conclude that, in the future, researchers who develop powerful evasion attacks should  evaluate their attacks against our region-based classification instead of standard point-based classification. 

In summary, our contributions are as follows:
\begin{itemize}

\item  \vspace{2mm} We perform a measurement study to characterize adversarial examples.

\item \vspace{2mm} We propose a region-based classification to defend against evasion attacks, while not impacting classification accuracy on benign examples.

\item \vspace{2mm} We evaluate our region-based classification using two image datasets. Our results demonstrate that 1) our method does not impact classification accuracy on benign examples, 2) our method is significantly more robust to existing evasion attacks than existing methods, and 3) our method is robust to the attacks that are strategically adjusted to our region-based classification. 

\end{itemize}


\section{Background and Related Work}
\label{relatedwork}

\subsection{Deep Neural Networks (DNNs)}
A deep neural network (DNN) consists of an input layer, several hidden layers, and an output layer. The output layer is often a \emph{softmax} layer. 
The neurons in one layer are connected with neurons in the next layer with certain patterns, e.g., fully connected, convolution, or max pooling~\cite{Krizhevsky12ImageNet}. In the \emph{training phase}, the weights on the connections are often learnt via back-propagation with a training dataset. 
In the \emph{testing phase}, the DNN is used to predict labels for examples that are unseen in the training phase. 
Specifically, suppose we have $L$ classes, denoted as $\{1,2,\cdots,L\}$. Both the layer before the output layer and the output layer have $L$ neurons. 
Let ${x}\in R^n$ be an unseen example, which is a $n$-dimension vector; $x_j$ represents the $j$th dimension of $x$. 
We denote the output of the $i$th neuron before the output layer as  $Z_i({x})$, and 
we denote the output of the $i$th neuron in the output layer as $F_i({x})$, where $i=1,2,\cdots,L$.  The outputs $Z_1({x}), Z_2({x}), \cdots, Z_L({x})$ are also called \emph{logits}.
Since the output layer is a softmax layer, $F_i({x})$ represents the probability that ${x}$ has a label $i$; 
and the $L$ outputs sum to 1, i.e., $\sum_{i=1}^L F_i({x})=1$. 
The label of ${x}$ is predicted to be the one that has the largest probability, i.e., $C({x})=\argmax_i F_i({x})$, where $C({x})$ is the predicted label.

A classifier essentially can be viewed as a \emph{classification boundary} that divides the $n$-dimension space into $L$ \emph{class regions}, denoted as $R_1$, $R_2$, $\cdots$, $R_L$. Any data point in the region $R_i$ will be predicted to have label $i$ by the classifier.

\begin{table}[!t]\renewcommand{\arraystretch}{1.2}
\centering
\caption{ evasion attacks.}
\centering
\begin{tabular}{|c|c|c|} \hline 
{Attack} & Type & {Noise metric} \\ \hline
{T-FGSM~\cite{goodfellow2014explaining}} & {Targeted} & {$L_\infty$} \\ \hline
T-IGSM~\cite{IGS16} & Targeted & $L_\infty$\\ \hline
T-JSMA~\cite{Papernot16} & Targeted & $L_0$\\ \hline
T-CW-$L_0$~\cite{CarliniSP17} &  {Targeted} &$L_0$\\ \hline
T-CW-$L_2$~\cite{CarliniSP17} &   {Targeted} &$L_2$\\ \hline
T-CW-$L_\infty$~\cite{CarliniSP17} &   {Targeted} & $L_\infty$\\ \hline
{U-FGSM~\cite{goodfellow2014explaining}} & {Untargeted} & {$L_\infty$} \\ \hline
U-IGSM~\cite{IGS16} & Untargeted & $L_\infty$\\ \hline
U-JSMA~\cite{Papernot16} & Untargeted & $L_0$\\ \hline
U-CW-$L_0$~\cite{CarliniSP17}& Untargeted & $L_0$\\ \hline
U-CW-$L_2$~\cite{CarliniSP17}& Untargeted & $L_2$\\ \hline
U-CW-$L_\infty$~\cite{CarliniSP17}& Untargeted & $L_\infty$\\ \hline
DeepFool~\cite{deepfool} & Untargeted & $L_2$\\ \hline
\end{tabular}
\label{attack}
\end{table}

\subsection{Evasion Attacks}
\label{relatedattack}
\emph{Poisoning attacks} and \emph{evasion attacks}~\cite{huang2011adversarial} are two well-known attacks to machine learning/data mining.  A poisoning attack aims to pollute the training dataset such that the learner produces a bad classifier. Various studies have demonstrated poisoning attacks to spam filter~\cite{Nelson08poisoningattackSpamfilter}, support vector machines~\cite{biggio2012poisoning}, deep neural networks~\cite{ShenACSAC16}, and recommender systems~\cite{poisoningattackRecSys16,YangRecSys17}. 
In an evasion attack, an attacker adds a small noise to a normal testing example (we call it \emph{benign example}) such that a classifier predicts an incorrect label for the example with noise. A testing example with noise is called \emph{adversarial example}. 
From a perspective of geometrics, an evasion attack moves a testing example from one class region to another.  

In this work, we focus on DNNs and evasion attacks. Evasion attacks can be classified into two categories, i.e., \emph{targeted evasion attacks} and \emph{untargeted evasion attacks}. In a targeted evasion attack, an attacker aims to add noise to a benign example such that the classifier predicts a particular incorrect label for the example. In an untargeted evasion attack, an attacker aims to mislead the classifier to predict any incorrect label. Table~\ref{attack} shows representative evasion attacks to DNNs, where the attacks with a prefix ``T-'' are targeted evasion attacks and the attacks with a prefix ``U-" are untargeted evasion attacks.

\subsubsection{Targeted Evasion Attacks} 

We denote by $C$ a DNN classifier.  $C({x})$ is the predicted label of a testing example ${x}\in [0,1]^n$. Note that we assume each dimension of $x$ is normalized to be in the range $[0,1]$, like previous studies~\cite{deepfool,CarliniSP17}.
 Szegedy et al.~\cite{szegedy2013intriguing} formally defined targeted evasion attacks as solving the following optimization problem:
\begin{align}
\label{evasionAttack}
\min\ & d({x}, {x} + {\delta}) \nonumber \\
\text{subject to: } & C(x+\delta) = t \nonumber \\
			 & x+\delta \in [0,1]^n, 
\end{align}
where $\delta$ is the added noise, $t$ is the \emph{target label} that the attacker wants the classifier to predict for the adversarial example $x+\delta$, and $d$ is a metric to measure distance between the benign example and the adversarial example. The label $t$ is not the true label of $x$. $L_0$, $L_2$, and $L_\infty$ norms are often used as the distance metric $d$. Specifically, $L_0$ norm is the number of dimensions of $x$ that are changed, i.e., the number of non-zero dimensions of $\delta$; $L_2$ norm is the standard Euclidean distance between $x$ and $x+\delta$; and $L_\infty$ norm is the maximum change to any dimension of $x$, i.e., $\max\{\delta_1, \delta_2, \cdots, \delta_n\}$.   

An algorithm to solve the optimization problem in Equation~\ref{evasionAttack} is called a targeted evasion attack. An adversarial example is successful if the classifier predicts the target label $t$ for it. The \emph{success rate (SR)} of a targeted evasion attack is the fraction of adversarial examples generated by the attack that are successful.

\myparatight{Targeted Fast Gradient Sign Method (T-FGSM)~\cite{goodfellow2014explaining}} Goodfellow et al.~\cite{goodfellow2014explaining} proposed a targeted Fast Gradient Sign Method (T-FGSM) based on the hypothesis that the classification boundary of a DNN is piecewise linear. T-FGSM is designed to  generate adversarial examples fast, without necessarily minimizing the added noise. Therefore, the adversarial examples generated by T-FGSM often have lower success rates than other optimized attacks when adding small noise~\cite{CarliniSP17}.  Formally, given a benign example ${x}$, T-FGSM generates an adversarial example ${x}'$ as follows:
\begin{align}
{x}' = {x} - \epsilon \cdot \text{sign}(\nabla_x J(\theta, {x}, t)),
\end{align}
where $\theta$ is the model parameters of the DNN, $\nabla$ indicates gradient, $t$ is the target label, $\epsilon$ is a parameter to control tradeoffs between the added noise and success rate of T-FGSM, and $J$ is the cost function used to train the DNN. Note that T-FGSM aims to minimize the $L_\infty$ norm of the added noise. Like Carlini and Wagner~\cite{CarliniSP17}, we search over $\epsilon$ to find the smallest noise that generates a successful adversarial example in our experiments; failure is returned if no $\epsilon$ produces a successful adversarial example.

\myparatight{Targeted Iterative Gradient Sign Method (T-IGSM)~\cite{IGS16}} Kurakin et al.~\cite{IGS16} proposed a targeted Iterative Gradient Sign Method (T-IGSM), which is an advanced version of the targeted Fast Gradient Sign Method (T-FGSM)~\cite{goodfellow2014explaining}. 
 Roughly speaking, T-IGSM iteratively adds small noise to the benign example until finding a successful adversarial example or reaching the maximum number of iterations; 
 in each iteration, T-IGSM clips the current adversarial example to be in the $L_\infty$ $\epsilon$-neighborhood of the benign example. Formally, T-IGSM works as follows:
\begin{align}
{x}'_0={x},\  {x}'_{N+1}=Clip_{{x},\epsilon}({x}'_{N}-\alpha \cdot \text{sign}(\nabla_x J(\theta, {x}, t))),
\end{align}
where $\theta$ is the model parameters of the DNN classifier, $\nabla$ indicates gradient, $t$ is the target label, $\epsilon$ is a parameter to control tradeoffs between the added noise and success rate of T-IGSM,  $J$ is the cost function used to train the DNN, $\alpha$ is a small step size, and the function $Clip_{{x},\epsilon}$ clips the current adversarial example to be in the $L_\infty$ $\epsilon$-neighborhood of ${x}$.  T-IGSM also aims to minimize the $L_\infty$ norm of the added noise. Like Carlini and Wagner~\cite{CarliniSP17}, we fix $\alpha=\frac{1}{256}$ and search over $\epsilon$ to find the smallest noise that generates a successful adversarial example; failure is returned if no $\epsilon$ produces a successful adversarial example.

\myparatight{Targeted Jacobian-based Saliency Map Attack (T-JSMA)~\cite{Papernot16}} Papernot et al.~\cite{Papernot16} proposed a targeted Jacobian-based Saliency Map Attack (T-JSMA). The attack is optimized to find adversarial examples with small $L_0$-norm noise.  T-JSMA iteratively adds noise to a benign example until the classifier $C$ predicts the target label $t$ as its label or the maximum number of iterations is reached. 
In each iteration, T-JSMA picks one or two entries of the example, by modifying which the example is most likely to move towards the target label $t$, and then the attack increases or decreases the entries by a constant value. Selecting the entries is assisted by the Jacobian-based saliency map. 
T-JSMA has two variants. One variant picks the entries to be modified via the softmax outputs of the DNN, while the other picks the entries via the logits of the DNN. We adopt the latter one as suggested by its authors~\cite{Papernot16}. Note that defensive distillation can prevent the variant of JSMA that uses softmax output, but not the variant that uses logits.

\myparatight{T-CW-$L_2$ attack~\cite{CarliniSP17}} Carlini and Wagner~\cite{CarliniSP17}  proposed a family of targeted evasion attacks, which generate successful adversarial examples with small noise. For simplicity, we call their attacks targeted Carlini and Wagner (T-CW) attacks. T-CW attacks have three variants that are tailored to the $L_0$, $L_2$, and $L_\infty$ norms, respectively. 
The variant T-CW-$L_2$ attack is tailored to find adversarial examples with small noise measured by $L_2$ norm. 
Formally, the evasion attack reformulates the optimization problem in Equation~\ref{evasionAttack} as the following optimization problem:
\begin{align}
\label{CW-L2}
\min\ & ||\frac{1}{2}(\text{tanh}(w)+1)-x||_2^2 + c \times f(\frac{1}{2}(\text{tanh}(w)+1)),
\end{align}
where $f(x')=\max(\max\{Z_i(x'):i \neq t\}- Z_t(x'), -k)$. The adversarial example is $\frac{1}{2}(\text{tanh}(w)+1)$, which automatically constrains each dimension to be in the range [0,1]. The parameter $k$ controls the confidence of the attack. By default, we set $k=0$. The noise $\delta$ is $\delta=\frac{1}{2}(\text{tanh}(w)+1) - x$. T-CW-$L_2$ iterates over the parameter $c$ via binary search in a relatively large range of candidate values. For each given $c$,  T-CW-$L_2$ uses the Adam optimizer~\cite{Adam14} to solve the optimization problem in Equation~\ref{CW-L2} to find the noise. The iterative process is halted at the smallest parameter $c$ that the classifier predicts the target label $t$ for the adversary example.

\myparatight{T-CW-$L_0$ attack~\cite{CarliniSP17}} This variant is tailored to find adversarial examples with small noise measured by $L_0$ norm. This attack iteratively identifies the dimensions of $x$ that do not have much impact on the classifier's prediction and fixes them. The set of fixed dimensions increases until the attack has identified a minimum subset of dimensions that can be changed to construct a successful adversarial example. In each iteration, the set of dimensions that can be fixed are identified by the T-CW-$L_2$ attack.  Specifically, in each iteration, T-CW-$L_0$ calls T-CW-$L_2$, which can only modify the unfixed dimensions. Suppose $\delta$ is the found noise for the benign example $x$. T-CW-$L_0$ computes the gradient $g=\nabla f(x+\delta)$ and selects the dimension $i=\argmin_i g_i \times \delta_i$ to be fixed. The iterative process is repeated until  T-CW-$L_2$ cannot find a successful adversarial example. Again, the parameter in T-CW-$L_2$ is selected via a searching process: starting from a very small $c$ value; if T-CW-$L_2$ fails, then doubling $c$ until finding a successful adversarial example.

\myparatight{T-CW-$L_\infty$ attack~\cite{CarliniSP17}} This variant is tailored to find adversarial examples with small noise measured by $L_\infty$ norm. This attack transforms the optimization problem in Equation~\ref{evasionAttack} to the following one:
\begin{align}
\label{CW-Li}
\min \sum_i (\delta_i - \tau)^+ + c \times f(x + \delta),
\end{align}
where $f$ is the same function as in T-CW-$L_2$; $(\delta_i - \tau)^+=0$ if $\delta_i < \tau$, otherwise $(\delta_i - \tau)^+=\delta_i - \tau$. T-CW-$L_\infty$ iterates over $c$ until finding a successful adversarial example. Specifically, $c$ is iteratively doubled from a small value. For each given $c$, CW-$L_\infty$ further iterates over $\tau$.   
In particular, $\tau$ is initialized to be 1. For a given $\tau$, T-CW-$L_\infty$ solves the optimization problem in Equation~\ref{CW-Li}. If $\delta_i < \tau$ for every $i$, then $\tau$ is reduced by a factor of 0.9, and then T-CW-$L_\infty$ solves the optimization problem with the updated $\tau$. This process is repeated until such a noise vector $\delta$ that $\delta_i < \tau$ for every $i$ cannot be found.

\subsubsection{Untargeted Evasion Attacks} In an untargeted evasion attack, an attacker aims to solve the following optimization problem:
\begin{align}
\label{unevasionAttack}
\min\ & d({x}, {x} + {\delta}) \nonumber \\
\text{subject to: } & C(x+\delta) \neq C^*(x) \nonumber \\
			 & x+\delta \in [0,1]^n, 
\end{align}
where $\delta$ is the added noise, $C^*(x)$ is the true label of $x$, and $d$ is a metric to measure distance between the benign example and the adversarial example. An algorithm to solve the optimization problem in Equation~\ref{unevasionAttack} is called an untargeted evasion attack. An adversarial example is successful if the classifier predicts a label that does not equal $C^*(x)$ for it. 

\myparatight{U-FGSM~\cite{goodfellow2014explaining}, U-IGSM~\cite{IGS16}, U-CW-$L_0$~\cite{CarliniSP17}, U-CW-$L_2$~\cite{CarliniSP17}, and U-CW-$L_\infty$~\cite{CarliniSP17}}  
Carlini and Wagner~\cite{CarliniSP17} proposed a strategy to convert a targeted evasion attack to an untargeted evasion attack. 
Suppose we have a targeted evasion attack $A$. Given a benign example $x$, whose true label is $C^*(x)$, we use $A$ to generate an adversarial example for each target label $t$ that does not equal $C^*(x)$. The adversarial example with the smallest noise is treated as the untargeted adversarial example for the benign example $x$. We use this strategy to transform the targeted evasion attacks T-FGSM, T-IGSM, 
T-CW-$L_0$, T-CW-$L_2$, and T-CW-$L_\infty$ to untargeted evasion attacks, which we denote as U-FGSM, U-IGSM, and U-CW-$L_0$, U-CW-$L_2$, and U-CW-$L_\infty$, respectively. 
We note that Goodfellow et al.~\cite{goodfellow2014explaining} proposed an untargeted FGSM attack, which constructs an adversarial example as ${x}' = {x} + \epsilon \cdot \text{sign}(\nabla_x J(\theta, {x}, C^*(x)))$. Moreover, IGSM has an untargeted version, which iteratively constructs an adversarial example as ${x}'_0={x},\  {x}'_{N+1}=Clip_{{x},\epsilon}\{{x}'_{N} + \alpha \cdot \text{sign}(\nabla_x J(\theta, {x}, C^*(x)))\}$. However, we found such untargeted versions add larger noise. Therefore, we will not use them in our experiments.

\myparatight{DeepFool~\cite{deepfool}} Moosavi-Dezfooli et al.~\cite{deepfool} proposed an untargeted evasion attack called DeepFool to differentiable classifiers. The key idea of DeepFool is to iteratively add noise to a benign example until the classifier predicts an incorrect label for the example or the maximum number of iterations is reached. In each iteration, DeepFool linearizes the classifier at the current adversarial example and finds the minimum noise required to move the adversarial example to the linearized classification boundary.

\subsubsection{Evaluation Metrics}
\label{evaluationmetric}
 An adversarial example is successful if it satisfies two conditions: 1) the adversarial example and the original benign example have the same true label (determined by human) and 2) the classifier predicts the target label $t$ (targeted evasion attacks) or an incorrect label (untargeted evasion attacks) for the adversarial example. It is unclear how to check the first condition automatically because we do not have a way to model human perception yet. In principle, \emph{success rate} of an evasion attack should be the fraction of its generated adversarial examples that satisfy both conditions. However, 
due to the challenges of checking the first condition, existing studies approximate {success rate} of an attack as the fraction of its generated adversarial examples that satisfy the second condition alone. Moreover, they also use the noise (measured by $L_0$, $L_2$, or $L_\infty$ norms) in the adversarial examples to supplement the approximate success rate. Therefore, in this work, we will use the approximate success rate and noise to measure evasion attacks. An evasion attack with a larger approximate success rate and/or a smaller noise is better. For simplicity, we will use approximate success rate  and success rate interchangeably.

\subsection{Defenses Against Evasion Attacks} 

\subsubsection{Detecting Adversarial Examples} One line of research~\cite{detection2,detection1,gong2017adversarial,li2016adversarial,feinman2017detecting,hendrycks2017early,xu2017feature} aim to detect adversarial examples, i.e., distinguish between benign examples and adversarial examples. Essentially, detecting adversarial examples is to design another binary machine learning classifier, which classifies a testing example to be benign or adversarial. An attacker can strategically adjust its attacks to evade both the original classifier and the new classifier to detect adversarial examples. Carlini and Wagner~\cite{Carlini17} demonstrated that, for such adaptive attacks, some detectors are ineffective while some detectors enforce attackers to add larger noise to construct successful adversarial examples. A key limitation of detecting adversarial examples is that it is unclear how to handle the testing examples that are predicted to be adversarial examples. We suspect that those testing examples eventually would require human to manually label them, i.e., the entire system becomes a human-in-the-loop system. For real-time automated decision making systems such as self-driving cars, it is challenging to require human to manually label the detected adversarial examples.

Meng and Chen proposed MagNet~\cite{MagNet}, an approach combining detection and de-noising. 
Specifically, given a testing example, they first use a {detector} to determine whether the testing example is an adversarial example or not.  If the testing example is predicted to be an adversarial example, the DNN classifier will not predict its label. If the testing example is not predicted to be an adversarial example, they will reform the testing example using a \emph{reformer}, which essentially de-noises the testing example via an autoencoder~\cite{GuRigazio}.  In the end, the DNN classifier will predict label of the reformed testing example and treat it as the label of the original testing example. MagNet designs both the detector and the reformer using auto-encoders, which are trained using only benign examples. Meng and Chen demonstrated that MagNet can reduce the success rates of various known evasion attacks. However, 
MagNet has two key limitations. First, MagNet decreases the classification accuracy on benign testing examples. For instance, on the CIFAR-10 dataset, their trained point-based DNN achieves an accuracy of 90.6\%. However, MagNet reduces the accuracy to be 86.8\% using the same point-based DNN. Second, like all methods to detect adversarial examples, MagNet relies on manually labeling the detected adversarial examples, losing the benefits of automated decision making.

\subsubsection{Building Robust Classifiers} Another line of research aim to design new methods to train DNNs.

\myparatight{Adversarial training} Goodfellow et al.~\cite{goodfellow2014explaining} proposed to train a DNN via augmenting the training dataset with adversarial examples, which is called \emph{adversarial training}. Specifically, for each training benign example, the learner generates a training adversarial example using evasion attacks. Then, the learner uses a standard algorithm (e.g., back propagation) to learn a DNN using the original training benign examples and the adversarial examples. Several variants~\cite{DataGrad,aTScale,tramer2017ensemble,madry2017towards,ATRobustOptimization} of adversarial training were also proposed. Adversarial training essentially reformulates the objective function used to learn model parameters. For instance, Madry et al.~\cite{madry2017towards} and Sinha et al.~\cite{ATRobustOptimization} formulated adversarial training as solving min-max optimization problems, which can be solved using robust optimization techniques. A key limitation of adversarial training is that it sacrifices classification accuracies on benign examples. For instance, in Madry et al.~\cite{madry2017towards}, the DNN classifier without adversarial training achieves a classification accuracy of 95.2\% on CIFAR-10; with adversarial training, the classification accuracy drops to 87.3\%.

\myparatight{Defensive distillation} Papernot et al.~\cite{Papernot16Distillation} proposed a distillation based method to train a DNN. The DNN is first trained using a standard method. For each training example, the DNN produces a vector of confidence scores. The confidence scores are treated as the soft label for the training example. Given the soft labels and the training examples, the weights of the DNN are retrained. A parameter $T$ named \emph{distillation temperature} is used in softmax layer during both training sessions to control confidence scores. 
Carlini and Wagner~\cite{CarliniSP17} demonstrated that their CW attacks  can still achieve 100\% success rates for DNNs trained with distillation. Moreover, the noises added to the benign examples when generating adversarial examples are just slightly higher for distilled DNNs than those for undistilled DNNs. Our experimental results confirm such findings.


\section{Design Goals}

We aim to achieve the following two goals:

{\bf 1) Not sacrificing classification accuracy on testing benign examples.} Our first design goal is that the defense method should maintain the high accuracy of the DNN classifier on testing benign examples. Neural networks re-gained unprecedented attention in the past several years under the coat of ``deep learning".  
The major reason is that neural networks with multiple layers (i.e., DNN) achieve significantly better classification accuracy than other machine learning methods for a variety of artificial intelligence tasks such as computer vision, speech recognition, and natural language processing.  Therefore, our defense method should maintain such advantage of DNNs. 

{\bf 2) Increasing robustness.}  We aim to design a defense method that is robust to powerful evasion attacks. 
In particular, our new classifier should have better robustness than conventional DNN classifiers with respect to state-of-the-art evasion attacks, e.g., the CW attacks.
Suppose we have a classifier $C$. After deploying a certain defense method, we obtain another classifier $D$. Suppose we have an evasion attack. The success rate (SR) of the attack and the average noise of the successful adversarial examples for the classifiers $C$ and $D$ are denoted as ($SR_C$, $\delta_C$) and ($SR_D$, $\delta_D$), respectively. We say that the classifier $D$ is more robust than the classifier $C$ if  $\delta_D \geq \delta_C$ and $SR_D \leq SR_C$, where the equality does not hold simultaneously for the two inequalities. In other words, a defense method is said to be effective with respect to an evasion attack if the method at least increases the noises of the generated successful adversarial examples or decreases the success rates.

 We note that our goal is not to completely eliminate adversarial examples. Instead, our goal is to increase robustness without sacrificing classification accuracy on benign examples. None of the existing methods to build robust classifiers satisfy the two goals simultaneously.

\begin{figure*}[!t]
\centering
\subfloat[T-CW-$L_2$ attack]{\includegraphics[width=0.85 \textwidth]{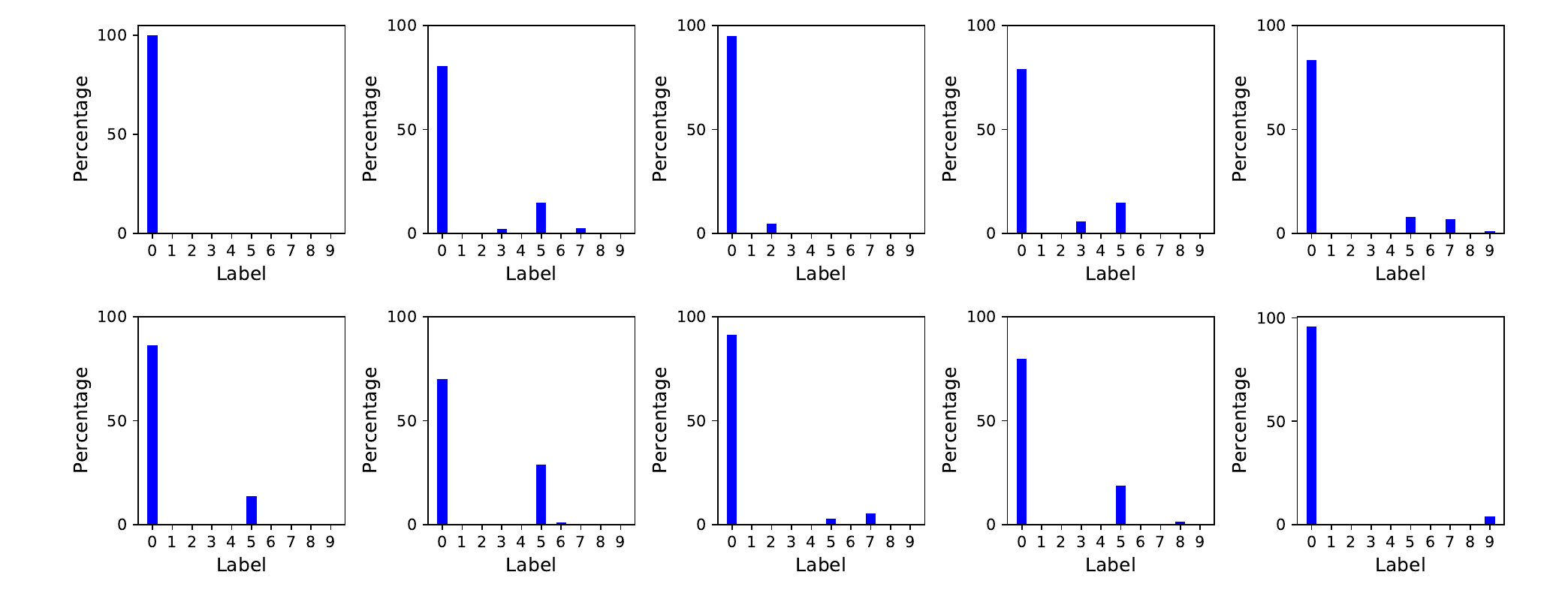}\label{l2}}\vspace{-2mm}

\subfloat[T-CW-$L_0$ attack]{\includegraphics[width=0.85 \textwidth]{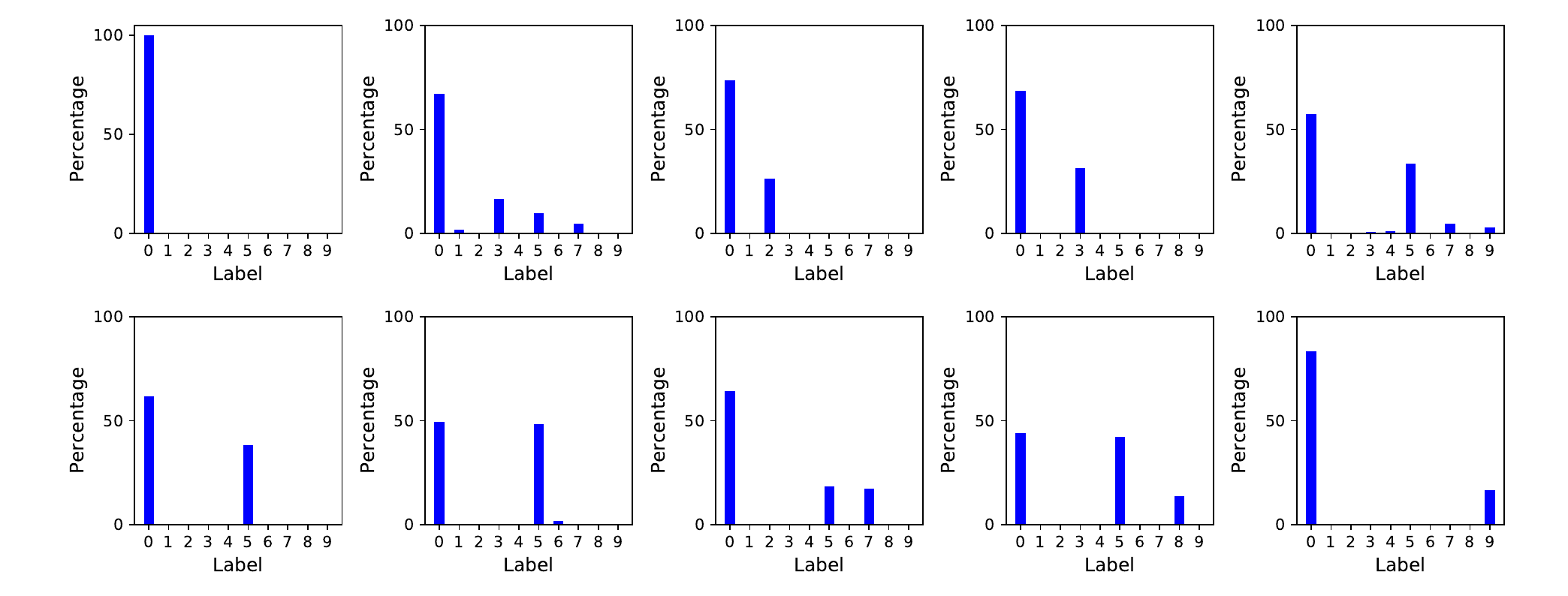}\label{l0}}\vspace{-2mm}

\subfloat[T-CW-$L_\infty$ attack]{\includegraphics[width=0.85 \textwidth]{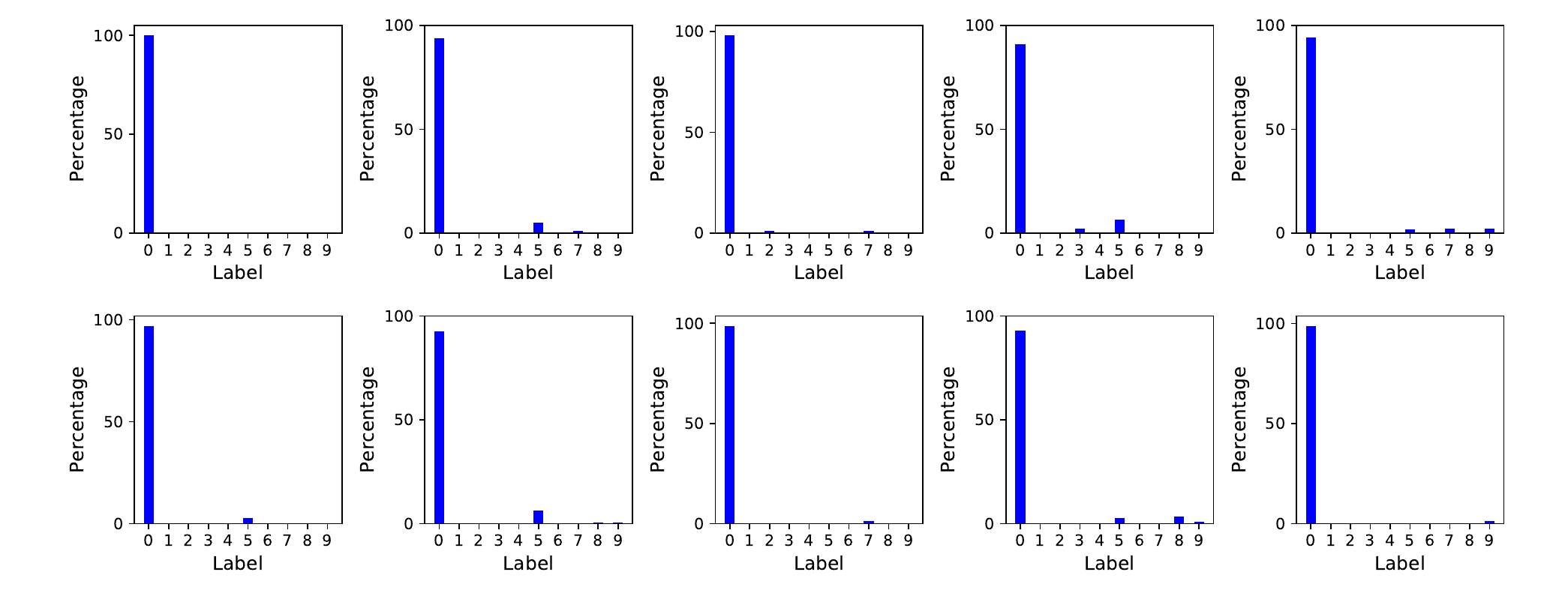}\label{li}}

\caption{Label histograms of 10,000 random data points in the hypercube around a benign example or its adversarial examples generated by the (a) T-CW-$L_2$ attack, (b) T-CW-$L_0$ attack, and (c) T-CW-$L_\infty$ attack. Each histogram corresponds to an example. The benign example has label 0. In each subfigure, the first row (from left to right): the benign example, and the adversarial examples that have target labels 1, 2, 3, and 4, respectively; and the second row (from left to right): the adversarial examples that have target labels 5, 6, 7, 8, and 9, respectively.}
\end{figure*}

\section{Measuring Evasion Attacks}
We first show some measurement results on evasion attacks, which motivate the design of our region-based classification method. 
We performed our measurements on the standard MNIST dataset. 
In the dataset, our task is to recognize the digit in an image, which is a 10-class classification problem. We normalize each pixel to be in the range [0,1].
We adopted the same DNN classifier that was used by Carlini and Wagner~\cite{CarliniSP17}. 
The classifier essentially classifies the digit image space into $10$ class regions, denoted as $R_0$, $R_1$, $\cdots$, $R_9$. Any data point in the class region $R_i$ will be predicted to have label $i$ by the classifier.

We sample a benign testing image of digit 0 uniformly at random. We use the T-CW-$L_2$, T-CW-$L_0$, and T-CW-$L_\infty$ attacks to generate adversarial examples based on the sampled benign example. We obtained the open-source implementation of the CW attacks from its authors~\cite{CarliniSP17}. For each target label $i$, we use an evasion attack to generate an adversarial example with the target label $i$ based on the benign example, where $i=1, 2, \cdots, 9$.  We denote the adversarial example with the target label $i$ as $x'(i)$. The DNN classifier predicts label $i$ for the adversarial example $x'(i)$, while its true label is 0.

We denote the \emph{hypercube} that is centered at $x$ and has a length of $r$ as $B(x, r)$. Formally,   $B(x, r)=\{y| y_j \in [0,1] \text{ and }  |y_j-x_j|\leq r, \forall j=1, 2, \cdots, n\}$, where $x_j$ and $y_j$ are the $j$th dimensions of $x$ and $y$, respectively. 
For each adversarial example $x'(i)$, 
we sample 10,000 data points from the hypercube $B(x'(i),r)$ uniformly at random, where we set $r=0.3$ in our experiments (we will explain the setting of $r$ in experiments). 
We treat each data point as a testing example and feed it to the DNN classifier, which predicts a label for it. 
For the 10,000 data points, we obtain a histogram of their labels predicted by the DNN classifier.

Figure~\ref{l2}, Figure~\ref{l0}, and Figure~\ref{li} show the label histograms for the 10,000 randomly sampled data points from the hypercube around the benign example and the 9 adversarial examples generated by the T-CW-$L_2$ attack, T-CW-$L_0$ attack, and  T-CW-$L_\infty$ attack, respectively. For instance, in Figure~\ref{l2}, the first graph in the first row shows the histogram of labels for the 10,000 data points that are sampled from the hypercube centered at the benign example; the second graph (from left to right) in the first row shows the histogram of labels for the 10,000 data points that are sampled from the hypercube centered at the adversarial example that has a predicted label 1, where the adversarial example is generated by the T-CW-$L_2$ attack.

For the benign example, almost all the 10,000 randomly sampled data points are predicted to have label 0, which is the true label of the benign example. 
For most adversarial examples, a majority of the 10,000 randomly sampled data points are predicted to have label 0, which is the true label of the adversarial examples. From these measurement results, we have the following two observations:
\begin{itemize}

\item {\bf Observation I:} The hypercube $B(x, r)$ centered at a benign example $x$ intersects the most with the class region $R_i$, where $i$ is the true label of the benign example $x$.  This indicates that we can still correctly predict labels for benign examples by ensembling information in the hypercube. 

\item {\bf Observation II:}  For most adversarial examples, the hypercube $B(x', r)$ intersects the most with the class region $R_i$, where $i$ is the true label of the adversarial example $x'$. This indicates that we can also correctly predict labels for adversarial examples by ensembling information in the hypercube.

\end{itemize}
These measurement results motivate us to design our region-based classification, which we will introduce in the next section.

\begin{figure}[!t]
\centering
\includegraphics[width=0.35 \textwidth]{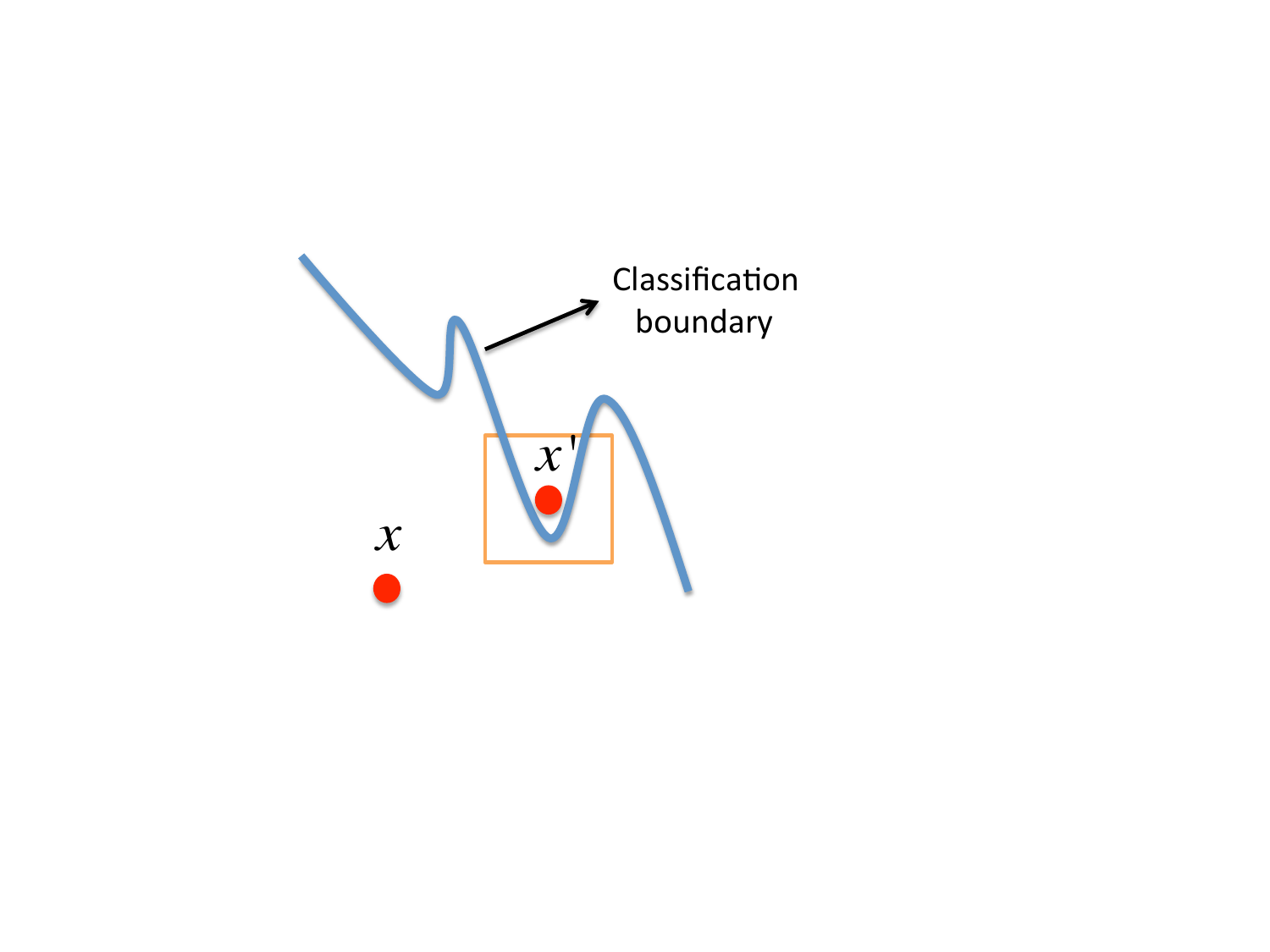}
\caption{Illustration of our region-based classification. $x$ is a testing benign example and $x'$ is the corresponding adversarial example. The hypercube centered at $x'$ intersects the most with the class region that has the true label.}
\label{method}
\end{figure}

\section{Our Region-based Classification}
We propose a defense method called \emph{Region-based Classification} (RC).  
Traditional DNN classifier is \emph{point-based}, i.e., given a testing example, the DNN classifier predicts its label. 
Therefore, we call such a classifier Point-based Classification (PC).  
In our RC classification, given a testing example, we ensemble information in the region around the testing example to predict its label. 
For any point-based DNN classifier, our method can transform it to be a region-based classifier that is more robust to adversarial examples, while maintaining its accuracy on benign examples.

\subsection{Region-based Classification}
Suppose we have a point-based DNN classifier $C$. For a testing example $x$ (either benign example or adversarial example), we create a hypercube $B(x, r)$ around the testing example. Recall that the DNN classifier essentially divides the input space into $L$ class regions, denoted as $R_1$, $R_2$, $\cdots$, $R_L$; all data points in the class region $R_i$ are predicted to have label $i$ by the classifier, where $i=1, 2, \cdots, L$. In our RC classifier, we predict the label of a testing example $x$ to be the one whose class region intersects the most with the hypercube $B(x, r)$.
We denote our RC classifier as $RC_{C,r}$ since it relies on the point-based DNN classifier $C$ and the length $r$. 
We denote the area of the intersection between $R_i$ and $B(x, r)$ as $A_i(x, r)$. Then, our classifier predicts the label of $x$ to be  $RC_{C,r}(x)=\argmax_i A_i(x,r)$.  Figure~\ref{method} illustrates our region-based classification. 

\myparatight{Approximating the areas $A_i(x,r)$} One challenge of using our RC classifier is how to compute the areas $A_i(x,r)$, because the class regions might be very irregular. We address the challenge via sampling $m$ data points from the hypercube $B(x, r)$ uniformly at random and use them to approximate the areas $A_i(x,r)$. In particular, for each sampled data point, we use the point-based classifier $C$ to predict its label. We denote by $a_i(x,r)$ the number of  sampled data points that are predicted to have label $i$ by the classifier $C$. Then, our RC classifier predicts the label of $x$ as $RC_{C,r}(x)=\argmax_i a_i(x,r)$.

\myparatight{Learning the length $r$} Another challenge for our RC classifier is how to determine the length $r$ of the hypercube. $r$ is a critical parameter for our method RC (we will show the impact of $r$ on the effectiveness of RC in our experiments). Specifically, $r$ controls the tradeoff between robustness to adversarial examples and classification accuracy on benign examples. Suppose we want to classify an adversarial example $x'$, whose true label is $i$. On one hand, if the length of the hypercube $B(x',r)$ is too small, the hypercube will not intersect with the class region $R_i$, which means that our RC classifier will not be able to correctly classify the adversarial example. On the other hand, if the length is too large, the hypercube around a benign example will intersect with the incorrect class regions, which makes our method predict incorrect labels for benign examples. 

To address the challenge, we propose to learn the length $r$ using a validation dataset consisting of only benign examples. We do not use adversarial examples because the adversarial examples used by the attacker may not be accessible to the defender.
Our key idea is to select the maximal length $r$ such that the classification accuracy of our classifier $RC_{C, r}$ on the validation dataset is no smaller than that of the point-based classifier $C$. There are many choices of $r$, with which our classifier $RC_{C, r}$ has no smaller classification accuracy than the point-based classifier $C$. We propose to select the maximum one,  so an adversarial example needs a larger noise to move further away from the classification boundary of $C$ in order to evade $RC_{C, r}$. 

Specifically, we learn the radius through a search process. Suppose a point-based DNN classifier $C$ has classification accuracy $ACC$ on the validation dataset. 
We transform the classifier $C$ into a RC classifier. 
Initially, we set $r$ to be a small value. For each benign example in the validation dataset, we predict its label using our classifier $RC_{C, r}$. 
We compute the classification accuracy of $RC_{C, r}$ on the validation dataset. If the classification accuracy is no smaller than $ACC$, we increase the radius $r$ by a \emph{step size} $\epsilon$ and repeat the process. This search process is repeated until the classifier $RC_{C, r}$ achieves a classification accuracy on the validation dataset that is smaller than $ACC$. Algorithm~\ref{radius} shows the search process.

\begin{algorithm}[t]
\caption{Learning Length $r$ by Searching}
\begin{algorithmic}[1]
\REQUIRE Validation dataset $V$, point-based DNN classifier $C$, step size $\epsilon$, initial length $r_0$. \\
\ENSURE  Length $r$. \\
         \STATE Initialize $r=r_0$. \;
	\STATE $ACC$ = Accuracy of $C$ on $V$. \;
	\STATE $ACC_{RC}$ =  Accuracy of the $RC_{C, r}$ classifier on $V$.\;
	
	\WHILE { $ACC_{RC} \geq ACC$} \;
	\STATE $r = r + \epsilon$. \;
	\STATE $ACC_{RC}$ =  Accuracy of the $RC_{C, r}$ classifier on $V$.\;
	\ENDWHILE \;
	
	\RETURN $r - \epsilon$. \;

\end{algorithmic}
\label{radius}
\end{algorithm}

\subsection{Evasion Attacks to Our RC Classifier}
\label{sec:attack}
We consider a strong attacker who knows all the model parameters of our classifier $RC_{C, r}$. 
In particular, the attacker knows the architecture and parameters of the point-based DNN classifier $C$, the length $r$, and $m$, the number of data points sampled to approximate the areas. Our threat model is also known as the \emph{white-box setting}. 

\subsubsection{Existing evasion attacks} An attacker can use any attack shown in Table~\ref{attack} to find adversarial examples to evade our classifier $RC_{C, r}$. All these evasion attacks require the classifier to be differentiable, in order to propagate the gradient flow from the outputs to the inputs. However, our classifier $RC_{C, r}$ is \emph{non-differentiable}. Therefore, we consider an attacker generates adversarial examples based on the point-based classifier $C$, which is the key component of our classifier $RC_{C, r}$; and the attacker uses the adversarial examples to attack $RC_{C, r}$. This is also known as transferring adversarial examples from one classifier to another. 

\myparatight{Combined evasion attacks} An attacker can also combine existing evasion attacks. In particular, for a benign example, the attacker performs each existing evasion attack to find an adversarial example; then the attacker uses the successful adversarial example that has the smallest noise with respect to a certain noise metric (i.e., $L_0$, $L_\infty$, or $L_2$) as the final adversarial example; failure is returned if  no evasion attacks can find a successful adversarial example. Specifically, the targeted combined attack T-CA-$L_0$ combines evasion attacks T-JSMA and T-CW-$L_0$; the targeted combined attack T-CA-$L_\infty$ combines evasion attacks T-FGSM, T-IGSM, and T-CW-$L_\infty$; the targeted combined attack T-CA-$L_2$ combines evasion attacks T-CW-$L_2$ with different confidence parameter $k$ (we searched $k$ until 40);  the untargeted combined attack U-CA-$L_0$ combines evasion attacks U-JSMA and U-CW-$L_0$; the untargeted combined attack U-CA-$L_\infty$ combines evasion attacks U-FGSM, U-IGSM, and U-CW-$L_\infty$; and the untargeted combined attack U-CA-$L_2$ combines evasion attacks U-CW-$L_2$ with different confidence parameter $k$ and DeepFool.

We used the open-source implementation from the corresponding authors for CW attacks and JSMA attacks, while we implemented the FGSM, IGSM, and DeepFool attacks by ourselves.

\subsubsection{New evasion attacks} An attacker, who knows our region-based classification, can also strategically adjust its attacks. 
 Specifically, since our classifier ensembles information within a region, an attacker can first use an existing evasion attack to find an adversarial example based on the point-based classifier $C$ and then strategically add more noise to the adversarial example. The goal is to move the adversarial example further away from the classification boundary such that the hypercube centered at the adversarial example does not intersect or intersects less with the class region that has the true label of the adversarial example.

Specifically, suppose we have a benign example $x$. The attacker uses an existing evasion attack to find the corresponding adversarial example $x'$. The added noise is $\delta=x'-x$. Then, the attacker strategically constructs another adversarial example as $x''=x + (1 + \alpha) \delta$. Essentially, the attacker moves the adversarial example further along the direction of the current noise.  
Note that, we will clip the adversarial example $x''$ to be in the space $[0,1]^n$. Specifically, for each dimension $i$ of $x''$, we set $x_i''=0$ if $x_i'' < 0$, we  $x_i''=1$ if $x_i'' > 1$, and $x_i''$ keeps unchanged if $0< x_i'' < 1$.  The parameter $\alpha$ controls how much further to move the adversarial example away from the classification boundary.  For $L_2$ and $L_\infty$ norms, $\alpha$ is the increased fraction of noise. Specifically, suppose an existing evasion attack finds an adversarial example $x'$ with noise $\delta$, whose $L_2$ and $L_\infty$ norms are $||\delta||_2$ and $||\delta||_\infty$, respectively. Then, the adapted adversarial example $x''$ has noise $(1 + \alpha) \delta$,  whose $L_2$ and $L_\infty$ norms are $(1+ \alpha) ||\delta||_2$ and $(1+ \alpha) ||\delta||_\infty$, respectively. 
A larger $\alpha$ indicates a larger noise (for $L_2$ and $L_\infty$ norms) and a possibly larger success rate.

For convenience, for an evasion attack, we append the suffix \emph{-A} at the end of the attack's name to indicate the attack that is adapted to our classifier $RC_{C,r}$. For instance, T-CW-$L_0$-A means the adapted version of the attack T-CW-$L_0$. 
In our experiments, we will explore how $\alpha$ impacts the success rates of the adapted evasion attacks and noises added to the adversarial examples.


\section{Evaluations}
\label{exp}

\begin{table}[!t]\renewcommand{\arraystretch}{1.2}
\centering
\caption{Dataset statistics.}
\centering
\begin{tabular}{|c|c|c|c|} \hline 
{\small } & {\small Training} & {\small Validation} &{\small Testing} \\ \hline
{\small MNIST} & {\small 55,000} & {\small 5,000} & {\small 10,000}\\ \hline
{\small CIFAR-10} & {\small 45,000} & {\small 5,000} & {\small 10,000}\\ \hline
\end{tabular}
\label{dataset}
\end{table}

\subsection{Experimental Setup}

\myparatight{Datasets} We perform evaluations on two standard image datasets used to benchmark object recognition methods: MNIST and CIFAR-10. Table~\ref{dataset} shows the statistics of the datasets.  For each dataset, we sample 5,000 of the predefined training examples uniformly at random and treat them as the validation dataset used to learn the length $r$ in our RC classifier.

\myparatight{Compared methods} We compare the following DNN classifiers.

\begin{figure}[!t]
\centering
\includegraphics[width=0.45 \textwidth]{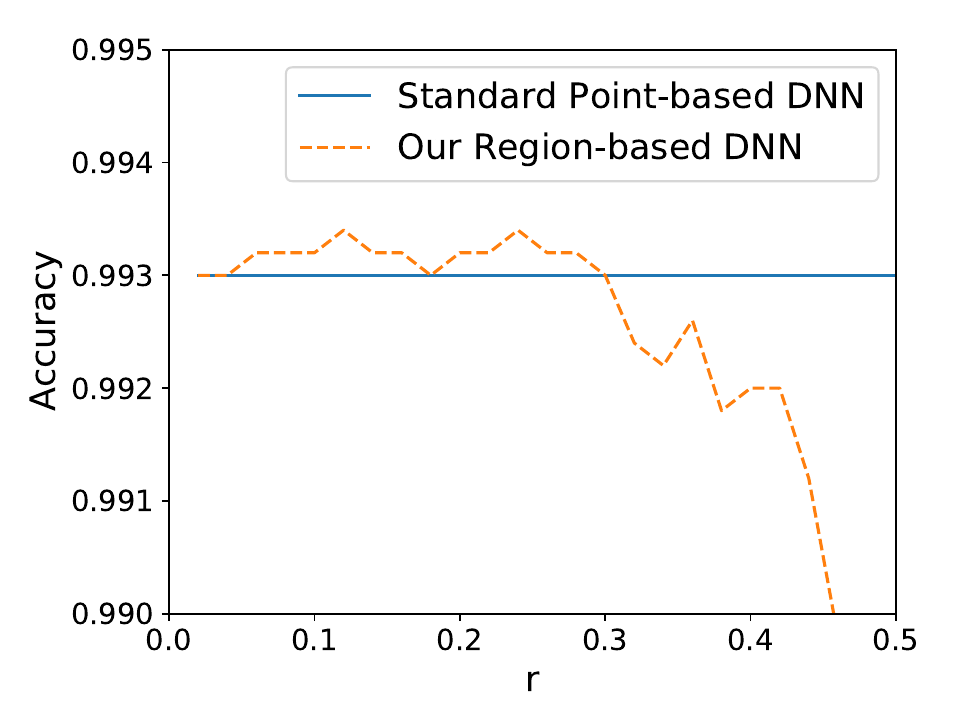}
\caption{Classification accuracies of the standard point-based DNN and our region-based DNN on the MNIST validation dataset as we increase the length $r$.}
\label{r}
\end{figure}

\begin{itemize}

\item {\bf Standard point-based DNN.} For each dataset, we trained a standard point-based DNN classifier. For the MNIST dataset, we adopt the same DNN architecture as the one adopted by Carlini and Wagner~\cite{CarliniSP17}. For the CIFAR-10 dataset, the DNN architecture adopted by Carlini and Wagner is not state-of-the-art.  Therefore, we do not adopt their DNN architecture for the CIFAR-10 dataset. Instead, we use the DNN architecture proposed by He et al.~\cite{CIFAR}. We obtained implementation from Carlini and Wagner to train the DNN for MNIST; and we obtained the implementation from~\cite{CIFAR10Implementation} to train the DNN for CIFAR-10.  

\item {\bf Adversarial training DNN.} For each dataset, we use adversarial training~\cite{goodfellow2014explaining} to learn a DNN classifier. The DNN classifiers have the same architectures as the standard point-based DNNs. 
The state-of-the-art adversarial training method was recently proposed by Madry et al.~\cite{madry2017towards}, which leverages robust optimization techniques. However, such adversarial training significantly sacrifices classification accuracy for benign examples. 
Therefore, we use the original adversarial training method proposed by Goodfellow et al.~\cite{goodfellow2014explaining} as a baseline robust classifier. Specifically, we use an evasion attack to generate adversarial example for each training example; and we use both the original training examples and the generated adversarial examples to train the DNN classifiers.  The evasion attack should have a high success rate, add small noise to adversarial examples, and be efficient. Considering the tradeoff between success rate, noise, and efficiency, we adopt DeepFool to generate adversarial examples in adversarial training.

\item {\bf Distillation DNN.} For each standard point-based DNN classifier, we use distillation~\cite{Papernot16Distillation} to re-train the DNN classifier with a temperature $T=100$. 

\item {\bf Our region-based DNN.} For each dataset, we transform the corresponding standard point-based DNN classifier to our region-based DNN classifier. The length $r$ is learnt through our Algorithm~\ref{radius} using the validation dataset. Specifically, we set the initial length value $r_0$ and step size $\epsilon$ in Algorithm~\ref{radius} to be 0 and 0.01, respectively. Figure~\ref{r} shows the classification accuracy of our RC classifier on the MNIST validation dataset as we increase the length $r$ in Algorithm~\ref{radius}. We observe that our classifier $RC_{C,r}$ has slightly higher accuracies than the standard point-based classifier $C$ when $r$ is small. Moreover, when $r$ is larger than around 0.3,  accuracy of $RC_{C,r}$ starts to decrease. Therefore, according to Algorithm~\ref{radius}, the length $r$ is set to be 0.3 for the MNIST dataset. Moreover, via Algorithm~\ref{radius}, the length $r$ is set to be 0.02 for the CIFAR-10 dataset. To estimate the areas between a hypercube and class regions, we sample 1,000 data points from the hypercube, i.e., the parameter $m$ is set to be 1,000.

\end{itemize}

\subsection{Classification Accuracies} 
Table~\ref{accuracy} shows the classification accuracies on testing benign examples of the compared classifiers for the MNIST and CIFAR-10 datasets. First, our region-based DNN achieves the same classification accuracy on the testing dataset with the standard point-based DNN for both the MNIST and CIFAR-10 datasets. This is because our region-based classifiers enable us to tune the length of the hypercube such that we do not sacrifice classification accuracies on testing benign examples. Second, adversarial training DNN and distillation DNN achieve lower classification accuracies than standard point-based DNN, though the differences are smaller for the MNIST dataset.  
In other words, adversarial training and distillation sacrifice classification accuracies for robustness.

\begin{table}[!t]\renewcommand{\arraystretch}{1.1}
\centering
\caption{Classification accuracy on benign testing examples. }
\centering
\begin{tabular}{|c|c|c|} \hline 
{ } & { MNIST} & CIFAR-10  \\ \hline
Point-based  & 99.4\% & 90.1\%  \\ \hline
Adversarial training & 99.3\%  & 88.1\% \\ \hline
Distillation & 99.2\%   & 88.3\% \\ \hline
Our region-based & 99.4\% & 90.1\%  \\ \hline
\end{tabular}
\label{accuracy}
\end{table}

\begin{table*}[!t]\renewcommand{\arraystretch}{1.1}
\centering
\caption{Success rates and average noise of successful adversarial examples for existing targeted evasion attacks to standard point-based and our region-based DNN classifiers.}
\centering

\subfloat[MNIST]{

\addtolength{\tabcolsep}{-4pt}
\begin{tabular}{|c|c|c|c|c|c|c|c|c|c|c|c|c|} \hline 

 & \multicolumn{4}{c|}{$L_0$} &  \multicolumn{6}{c|}{$L_\infty$}   &  \multicolumn{2}{c|}{$L_2$} \\  \cline{2-13}
  & \multicolumn{2}{c|}{T-JSMA} & \multicolumn{2}{c|}{T-CW-$L_0$} &  \multicolumn{2}{c|}{T-FGSM}  &  \multicolumn{2}{c|}{T-IGSM}   &  \multicolumn{2}{c|}{T-CW-$L_\infty$}   &  \multicolumn{2}{c|}{T-CW-$L_2$} \\  \cline{2-13}

& SR & Noise & SR & Noise & SR & Noise & SR & Noise & SR & Noise & SR & Noise \\ \hline

Point-based & 100\% & 72.3 & 100\% & 18.8 & 38\% & 0.276 & 99.9\% & 0.183 & 100\% & 0.188 & 100\% & 2.01 \\ \hline

Adversarial training & 100\% & 108.5 & 100\% & 17.5 & 45\% & 0.257 & 99.9\% & 0.139 & 100\% & 0.143 & 100\% & 1.41 \\ \hline

Distillation & 98\% & 45.3 & 100\% & 21.0 & 52\% & 0.232 & 100\% & 0.162 & 100\% & 0.163 & 100\% & 1.96 \\ \hline

Region-based & 53\% & 53.1 & 19.1\% & 11.9 & 11\% & 0.339 & 0.1\% & 0.086 & 0.1\% & 0.089 & 0.2\% & 0.912 \\ \hline

\end{tabular}
\label{targeted-MNIST}}

\subfloat[CIFAR-10]{

\addtolength{\tabcolsep}{-4pt}
\begin{tabular}{|c|c|c|c|c|c|c|c|c|c|c|c|c|} \hline 

 & \multicolumn{4}{c|}{$L_0$} &  \multicolumn{6}{c|}{$L_\infty$}   &  \multicolumn{2}{c|}{$L_2$} \\  \cline{2-13}
  & \multicolumn{2}{c|}{T-JSMA} & \multicolumn{2}{c|}{T-CW-$L_0$} &  \multicolumn{2}{c|}{T-FGSM}  &  \multicolumn{2}{c|}{T-IGSM}   &  \multicolumn{2}{c|}{T-CW-$L_\infty$}   &  \multicolumn{2}{c|}{T-CW-$L_2$} \\  \cline{2-13}

& SR & Noise & SR & Noise & SR & Noise & SR & Noise & SR & Noise & SR & Noise \\ \hline

Point-based & 100\% & 79.7 & 100\% & 30.5 & 72\% & 0.024 & 100\% & 0.008 & 100\% & 0.007 & 100\% & 0.192 \\ \hline

Adversarial training & 100\% & 84.3 & 100\% & 28.9 & 70\% & 0.056 & 100\% & 0.008 & 100\% & 0.01 & 100\% & 0.215 \\ \hline

Distillation & 100\% & 105.2 & 100\% & 32.6 & 64\% & 0.027 & 100\% & 0.009 & 100\% & 0.011 & 100\% & 0.251 \\ \hline

Region-based & 78\% & 85.6 & 6.3\% & 22.9 & 50\% & 0.025 & 29\% & 0.007 & 2.7\% & 0.004 & 2.6\% & 0.079 \\ \hline

\end{tabular}
\label{targeted-CIFAR}}

\label{targeted}
\end{table*}

\subsection{Robustness to Existing Evasion Attacks} 

We analyze robustness of our region-based classifiers with respect to existing targeted evasion attacks, untargeted evasion attacks, and ensemble evasion attacks.

\subsubsection{Targeted Evasion Attacks} Table~\ref{targeted} shows the success rates and average noise of successful adversarial examples for existing targeted evasion attacks.  Since the CW attacks are inefficient, for each dataset, we randomly sample 100 testing benign examples that the standard point-based DNN correctly classifies and generate adversarial examples for them. For each testing benign example and each targeted evasion attack, we generate an adversarial example for each candidate target label. More specifically, for each testing benign example and for each targeted evasion attack, we generate 9 adversarial examples since the MNIST and CIFAR-10 are 10-class classification problems. We compute the success rates of attacks using all these adversarial examples and the average noise using successful adversarial examples.  Note that these results are slightly different from those that we reported in our conference paper~\cite{region17}. This is because we generated one adversarial example for a randomly selected target label for a testing benign example in our conference paper. 

First, for each targeted evasion attack, the success rate is significantly lower for our region-based DNN classifier than for standard point-based DNN classifier. 
In other words, for our region-based classifier, existing targeted evasion attacks can construct successful adversarial examples for less number of testing benign examples. 
Moreover, compared with point-based classifier, some attacks have larger noise while some have smaller noise for our region-based classifier. This indicates that, for a given attack, among the adversarial examples generated based on different benign examples, the $L_0$, $L_2$, or $L_\infty$ norm does not necessarily measure how likely our region-based classifier can correctly predict the labels of the adversarial examples.

Second, when an attack has a smaller noise for the point-based classifier, the attack has a lower success rate for the region-based classifier. For instance, among the $L_0$-norm attacks, both T-CW-$L_0$ and T-JSMA have success rates of 100\% for the point-based classifier, while they have average noise of 18.8 and 72.3 on MNIST, respectively. However, for our region-based classifier,  T-CW-$L_0$ and T-JSMA have success rates of 19\% and 53\% on MNIST, respectively. Likewise, among the $L_\infty$-norm attacks, T-FGSM has the largest noise for the point-based classifier  
and the largest success rates for our region-based classifier. In other words, when an attack is better (adding smaller noise to construct successful adversarial examples) for the point-based classifier, the attack is worse (achieving lower success rates) for the region-based classifier. We speculate the reason is that if two attacks generate two adversarial examples for a given benign example, the adversarial example with a larger noise is more likely to be further away from the classification boundary and evade our region-based classifier. 

State-of-the-art targeted evasion attacks (e.g., CW attacks) aim to find adversarial examples with minimum noise, i.e.,  they find adversarial examples via solving the optimization problem in Equation~\ref{evasionAttack}. Such attacks can find successful adversarial examples with small noise for the standard point-based classifiers. Our observation indicates that such attacks have low success rates for the region-based classifiers. This is because the adversarial examples are close to classification boundary. To evade our region-based classifier, we may need to reformulate the optimization problem of finding adversarial examples.

Third, for our region-based classifiers, $L_0$-norm attacks can achieve the highest success rates among existing attacks. The reason is that the adversarial examples generated by the $L_0$-norm attacks could be further away from the classification boundary. However, $L_0$-norm attacks add ``spots" on benign image examples, which may be easier for human to perceive.  

\begin{figure}[!ht]
\centering
{\includegraphics[width=0.45 \textwidth]{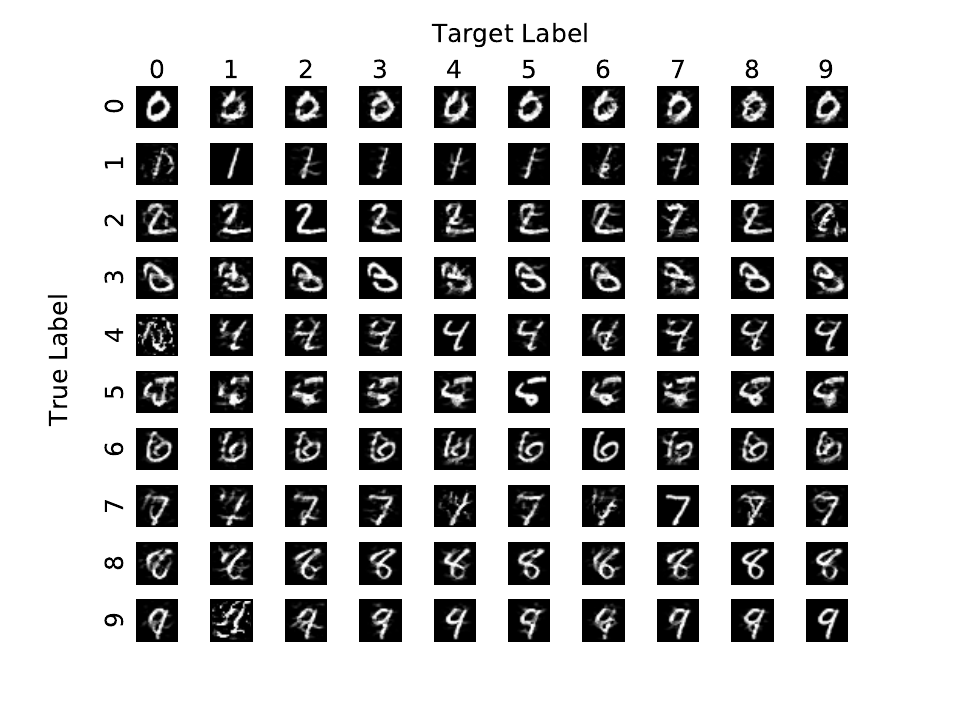}}
\vspace{-4mm}
\caption{Adversarial examples generated by the high-confidence T-CW-$L_2$ attack for 10 randomly selected benign examples, where $k=20$.}
\label{hcexample}
\end{figure}

\begin{figure*}[!t]
\centering
\subfloat[Success Rate, MNIST]{\includegraphics[width=0.45 \textwidth]{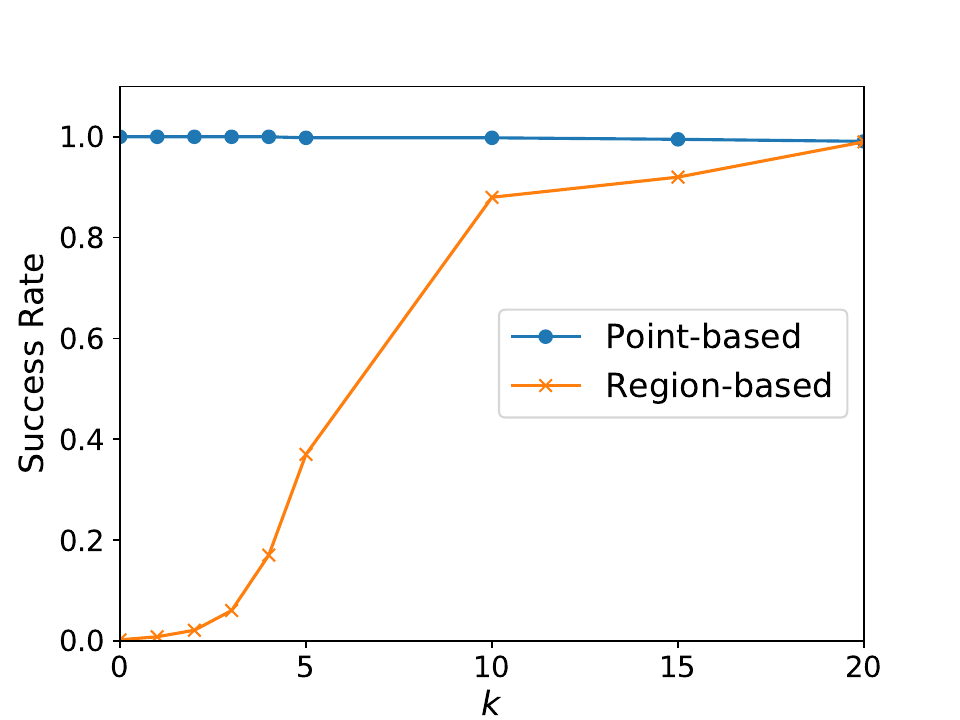}\label{mnist-sr}}
\subfloat[Noise, MNIST]{\includegraphics[width=0.45 \textwidth]{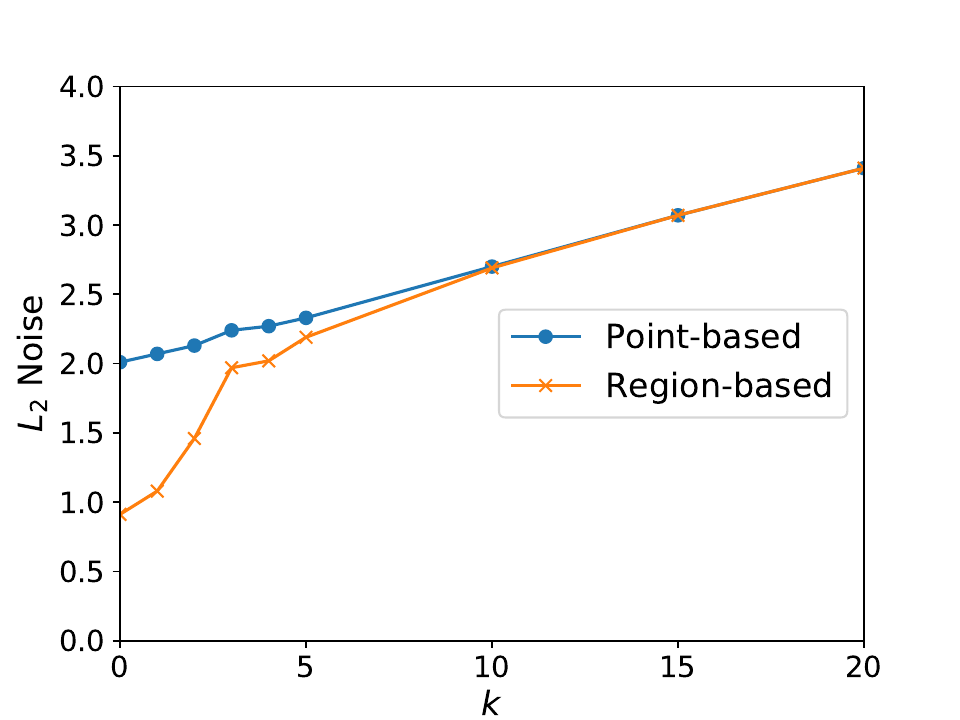}\label{mnist-noise}}

\vspace{-4mm}
\subfloat[Success Rate, CIFAR-10]{\includegraphics[width=0.45 \textwidth]{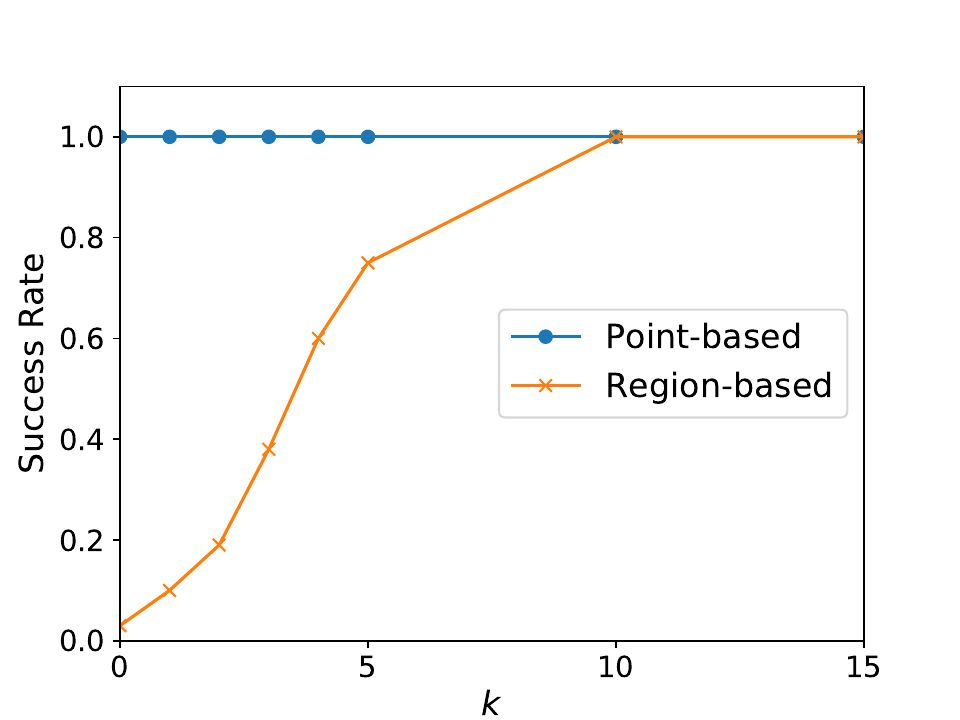}\label{cifar-sr}}
\subfloat[Noise, CIFAR-10]{\includegraphics[width=0.45 \textwidth]{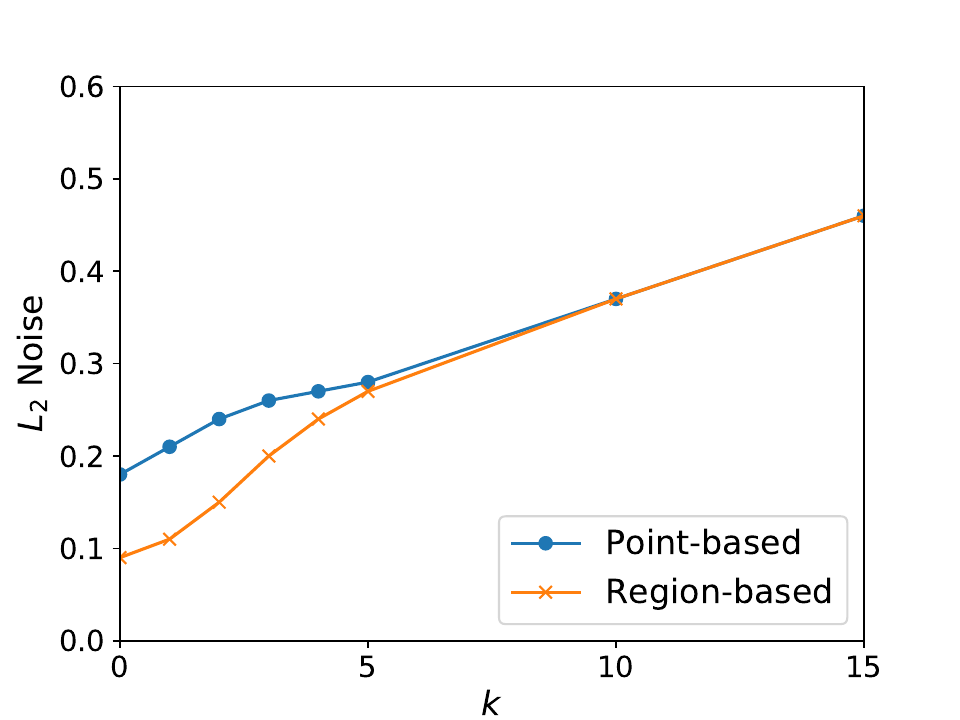}\label{cifar-noise}}
\caption{Success rates and average noise of successful adversarial examples for the high-confidence T-CW-$L_2$ attack.}
\vspace{-2mm}
\label{hc}
\end{figure*}

\subsubsection{High-confidence T-CW-$L_2$ Attack} Carlini and Wagner~\cite{CarliniSP17} also proposed a high-confidence version of T-CW-$L_2$ attacks, where the confidence is controlled by the parameter $k$ in Equation~\ref{CW-L2}. We use the high-confidence T-CW-$L_2$ attack to generate adversarial examples based on the point-based classifier, since our region-based classifier is non-differentiable.  Again, for each of the sampled testing benign example, we generate an adversarial examples for each candidate target label. Figure~\ref{hc} shows the success rates and average noise of successful adversarial examples for the high-confidence T-CW-$L_2$ attack, as we increase the confidence parameter $k$.

For our region-based classifier, the high-confidence T-CW-$L_2$ attack has a higher success rate with a larger $k$, but the noise is also larger. In particular, for the MNIST dataset, the high-confidence T-CW-$L_2$ attack has a success rate of 100\% when $k$ is around 20. However, when $k$ is around 20, the average noise is 70\% larger than that when $k=0$. In other words, to achieve a 100\% success rate, an attacker needs to add 70\% more noise on average for our region-based classifier than for the point-based classifier. Likewise, for the CIFAR-10 dataset, to achieve a 100\% success rate, an attacker needs to add 100\% more noise on average for our region-based classifier than for the point-based classifier. We note that our region-based classifier obtains such robustness gains without sacrificing classification accuracy on benign examples at all.

Figure~\ref{hcexample} shows adversarial examples generated by the high-confidence T-CW-$L_2$ attack for 10 randomly selected benign examples, where $k=20$. The examples on the diagonal are the benign examples; the examples on the $i$th row are supposed to have the true label of $i$; and the examples on the $j$th column are predicted to have a label $j$ by the region-based classifier.  However, a significant number of adversarial examples have changed the true label and are hard for human to recognize, e.g., benign example 2 with a target label 9, benign example 4 with a target label 0, and benign example 9 with a target label 1. Recall that in Section~\ref{evaluationmetric}, we discussed that a successful adversarial example should satisfy two conditions and we approximate success rate of an attack using its generated adversarial examples that satisfy the second condition only. Our results show that some adversarial examples that satisfy the second condition do not satisfy the first condition. Therefore, the real success rates of the high-confidence T-CW-$L_2$ attacks are lower than what we reported in Figure~\ref{hc}.

\begin{table*}[!t]\renewcommand{\arraystretch}{1.1}
\centering
\caption{Success rates and average noise of successful adversarial examples for existing untargeted evasion attacks to standard point-based and our region-based DNN classifiers. }
\centering

\subfloat[MNIST]{

\addtolength{\tabcolsep}{-4pt}
\begin{tabular}{|c|c|c|c|c|c|c|c|c|c|c|c|c|c|c|} \hline 

 & \multicolumn{4}{c|}{$L_0$} &  \multicolumn{6}{c|}{$L_\infty$}   &  \multicolumn{4}{c|}{$L_2$} \\  \cline{2-15}
  & \multicolumn{2}{c|}{U-JSMA} & \multicolumn{2}{c|}{U-CW-$L_0$} &  \multicolumn{2}{c|}{U-FGSM}  &  \multicolumn{2}{c|}{U-IGSM}   &  \multicolumn{2}{c|}{U-CW-$L_\infty$}   &  \multicolumn{2}{c|}{U-CW-$L_2$} &  \multicolumn{2}{c|}{DeepFool}  \\  \cline{2-15}

& SR & Noise & SR & Noise & SR & Noise & SR & Noise & SR & Noise & SR & Noise & SR & Noise \\ \hline

Point-based & 100\% & 21.4 & 100\% & 8.1 & 100\% & 0.188 & 100\% & 0.133 & 100\% & 0.138 & 100\% & 1.35 & 100\% &2.17 \\ \hline

Adversarial training & 100\% & 36.9 & 100\% & 8.2 & 100\% & 0.151 & 100\% & 0.098 & 100\% & 0.101 & 100\% & 0.932 & 100\% & 1.39 \\ \hline

Distillation & 100\% & 13.64 & 100\% & 9.2 & 100\% & 0.137 & 100\% & 0.107 & 100\% & 0.111 & 100\% & 1.26 & 100\% & 1.74 \\ \hline

Region-based & 98\% & 23.1 & 95\% & 11.8 & 80\% & 0.288 & 18\% & 0.222 & 12\% & 0.182 & 19\% & 2.42 & 34\% & 2.39 \\ \hline

\end{tabular}
\label{untargeted-MNIST}}

\subfloat[CIFAR-10]{

\addtolength{\tabcolsep}{-4pt}
\begin{tabular}{|c|c|c|c|c|c|c|c|c|c|c|c|c|c|c|} \hline 

 & \multicolumn{4}{c|}{$L_0$} &  \multicolumn{6}{c|}{$L_\infty$}   &  \multicolumn{4}{c|}{$L_2$} \\  \cline{2-15}
  & \multicolumn{2}{c|}{U-JSMA} & \multicolumn{2}{c|}{U-CW-$L_0$} &  \multicolumn{2}{c|}{U-FGSM}  &  \multicolumn{2}{c|}{U-IGSM}   &  \multicolumn{2}{c|}{U-CW-$L_\infty$}   &  \multicolumn{2}{c|}{U-CW-$L_2$} &  \multicolumn{2}{c|}{DeepFool}  \\  \cline{2-15}

& SR & Noise & SR & Noise & SR & Noise & SR & Noise & SR & Noise & SR & Noise & SR & Noise \\ \hline

Point-based & 100\% & 28.2 & 100\% & 12.5 & 100\% & 0.0065 & 100\% & 0.005 & 100\% & 0.010 & 100\% & 0.11 & 100\% & 0.16 \\ \hline

Adversarial training & 100\% & 24.9 & 100\% & 12.3 & 100\% & 0.007 & 100\% & 0.005 & 100\% & 0.006 & 100\% & 0.12 & 94\% & 0.19 \\ \hline

Distillation & 100\% & 39.1 & 100\% & 14.4 & 99\% & 0.0079 & 100\% & 0.006 & 100\% & 0.007 & 100\% & 0.136 & 84\% & 0.19 \\ \hline

Region-based & 100\% & 34.8 & 41\% & 20.8 & 100\% & 0.0079 & 85\% & 0.006 & 17\% & 0.009 & 16\% & 0.13 & 20\% & 0.15 \\ \hline

\end{tabular}
\label{untargeted-CIFAR}}

\label{untargeted}
\end{table*}

\begin{figure*}[!t]
\centering
\subfloat[Success Rate, MNIST]{\includegraphics[width=0.45 \textwidth]{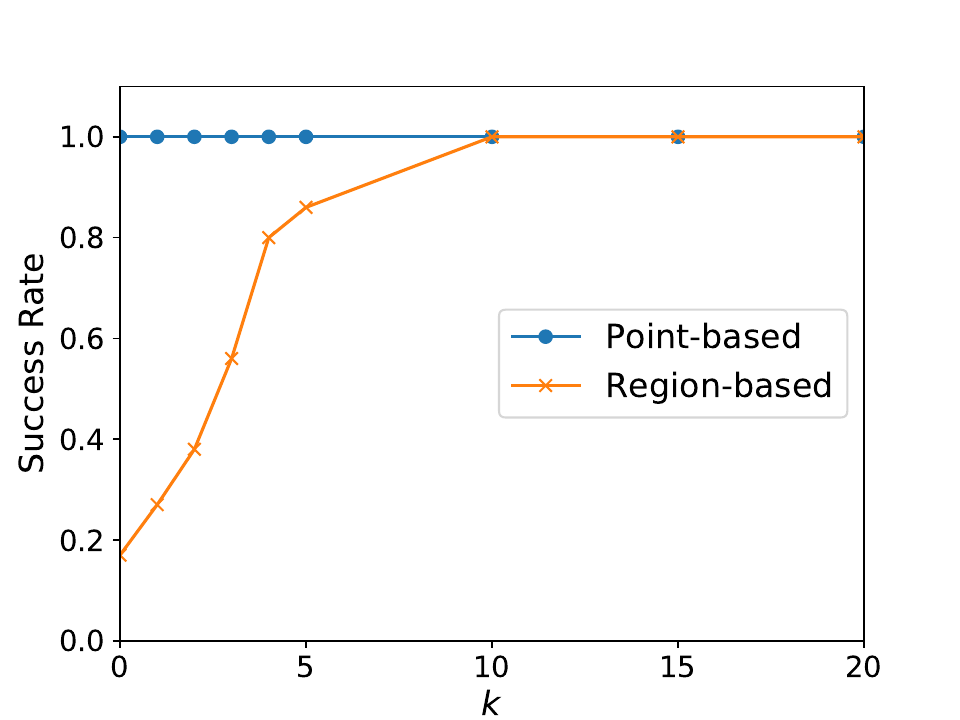}}
\subfloat[Noise, MNIST]{\includegraphics[width=0.45 \textwidth]{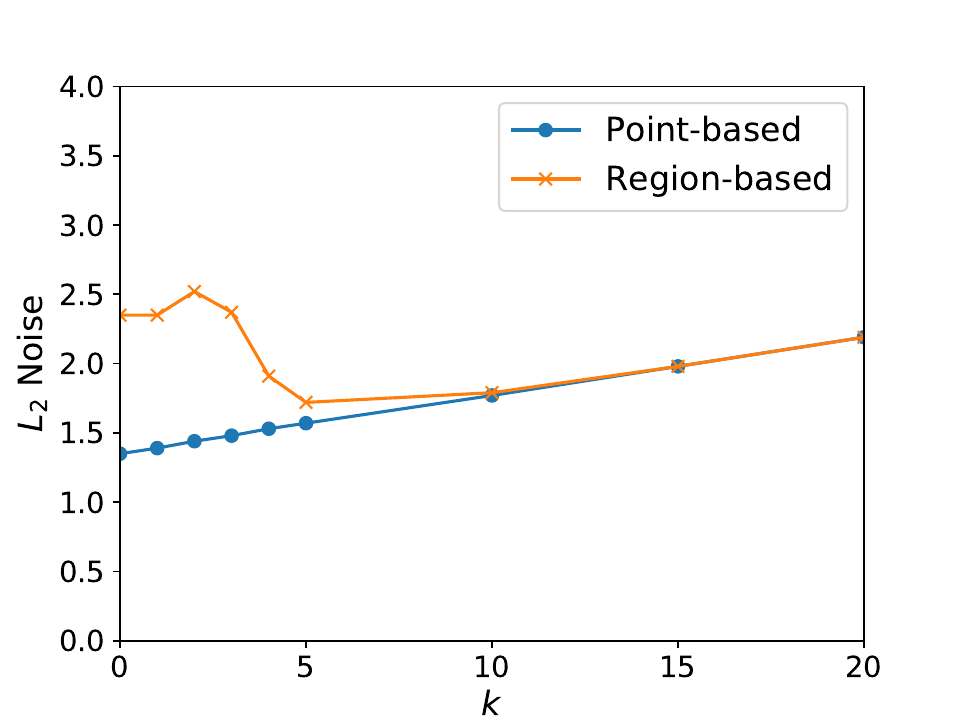}}

\subfloat[Success Rate, CIFAR-10]{\includegraphics[width=0.45 \textwidth]{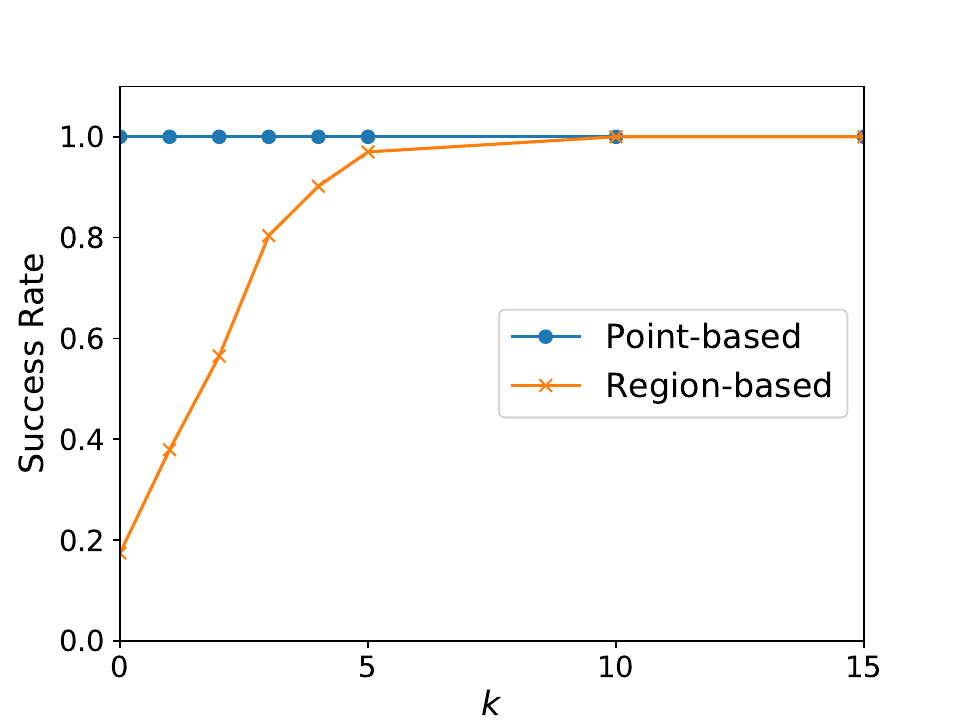}}
\subfloat[Noise, CIFAR-10]{\includegraphics[width=0.45 \textwidth]{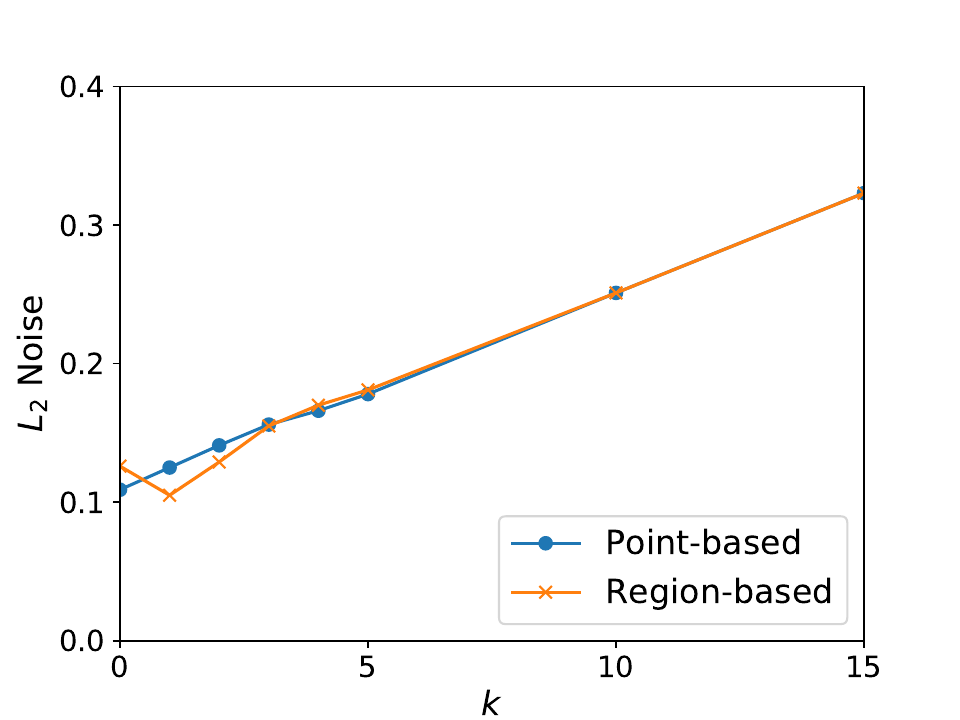}}
\caption{Success rates and average noise of successful adversarial examples for the high-confidence U-CW-$L_2$ attack.}
\label{unhc}
\end{figure*}

\begin{table*}[!t]\renewcommand{\arraystretch}{1.1}
\centering
\caption{Success rates and average noise of successful adversarial examples for combined evasion attacks to the compared DNN classifiers. }
\centering
\subfloat[Targeted combined attacks]{
\addtolength{\tabcolsep}{-4pt}
\begin{tabular}{|c|c|c|c|c|c|c|c|c|c|c|c|c|} \hline 

&  \multicolumn{6}{c|}{MNIST} &  \multicolumn{6}{c|}{CIFAR-10} \\ \cline{2-13}

& \multicolumn{2}{c|}{T-CA-$L_0$} & \multicolumn{2}{c|}{T-CA-$L_\infty$} & \multicolumn{2}{c|}{T-CA-$L_2$} & \multicolumn{2}{c|}{T-CA-$L_0$} & \multicolumn{2}{c|}{T-CA-$L_\infty$} & \multicolumn{2}{c|}{T-CA-$L_2$} \\ \cline{2-13}

& SR & Noise & SR & Noise & SR & Noise & SR & Noise & SR & Noise & SR & Noise \\ \hline

Point-based & 100\% & 18.8 & 100\% & 0.18 & 100\% & 2.01 & 100\% & 29.9 & 100\% & 0.007 & 100\% & 0.18 \\ \hline

Adversarial training & 100\% & 17.5 & 100\% & 0.14 & 100\% & 1.41 & 100\% & 29.7 & 100\% & 0.009 & 100\% & 0.22 \\ \hline

Distillation & 100\% & 21.0 & 100\% & 0.16 & 100\% & 1.95 & 100\% & 34.9 & 100\% & 0.009 & 100\% & 0.25 \\ \hline

Region-based & 55\% & 36.1 & 11\% & 0.34 & 100\% & 2.65 & 76\% & 85.5 & 58\% & 0.019 & 100\% & 0.26 \\ \hline

\end{tabular}
\label{ensemble-t}}

\subfloat[Untargeted combined attacks]{
\addtolength{\tabcolsep}{-4pt}
\begin{tabular}{|c|c|c|c|c|c|c|c|c|c|c|c|c|} \hline 

&  \multicolumn{6}{c|}{MNIST} &  \multicolumn{6}{c|}{CIFAR-10} \\ \cline{2-13}

& \multicolumn{2}{c|}{U-CA-$L_0$} & \multicolumn{2}{c|}{U-CA-$L_\infty$} & \multicolumn{2}{c|}{U-CA-$L_2$} & \multicolumn{2}{c|}{U-CA-$L_0$} & \multicolumn{2}{c|}{U-CA-$L_\infty$} & \multicolumn{2}{c|}{U-CA-$L_2$} \\ \cline{2-13}

& SR & Noise & SR & Noise & SR & Noise & SR & Noise & SR & Noise & SR & Noise \\ \hline

Point-based & 100\% & 8.0 & 100\% & 0.13 & 100\% & 1.35 & 100\% & 11.9 & 100\% & 0.005 & 100\% & 0.11 \\ \hline

Adversarial training & 100\% & 8.2 & 100\% & 0.10 & 100\% & 0.93 & 100\% & 12.8 & 100\% & 0.005 & 100\% & 0.12 \\ \hline

Distillation & 100\% & 9.2 & 100\% & 0.11 & 100\% & 1.26 & 100\% & 15.2 & 100\% & 0.006 & 100\% & 0.14 \\ \hline

Region-based & 100\% & 11.8 & 84\% & 0.28 & 100\% & 1.58 & 100\% & 29.3 & 100\% & 0.007 & 100\% & 0.19 \\ \hline

\end{tabular}
\label{ensemble-u}}

\label{ensemble}
\end{table*}

\subsubsection{Untargeted Evasion Attacks} Table~\ref{untargeted} shows the success rates and average noise of successful adversarial examples for existing untargeted evasion attacks, where the testing benign examples are the same as those used for evaluating the targeted evasion attacks. For a testing benign example, DeepFool directly finds an adversarial example that the classifier predicts an incorrect label. For each remaining untargeted evasion attack (e.g., U-JSMA), we have used its corresponding targeted version (e.g., T-JSMA) to generate 9 adversarial examples for each candidate target label in our experiments of evaluating targeted evasion attacks; we use the successful adversarial example that has the smallest noise as the adversarial example generated by the untargeted evasion attack; failure is returned if none of the 9 adversarial examples is successful.

Similar to targeted evasion attacks, every untargeted evasion attack has a lower success rate and/or larger average noise for our region-based classifier than for point-based classifier. Moreover, when an untargeted evasion attack is better (adding larger noise to construct successful adversarial examples) for the point-based classifier, the attack is very likely to be worse (achieving lower success rates) for the region-based classifier. Again, the reason is that adversarial examples that have larger noise are further away from the classification boundary and thus are more likely to evade our region-based classifiers. 
Compared to targeted evasion attacks, the corresponding untargeted versions have higher success rates and lower average noise. This is because an untargeted attack tries every candidate target label for a testing benign example; the attack is successful if the attack finds a successful adversarial example for at least one target label; and the successful adversarial example with the smallest noise is used as the adversarial example for the untargeted attack.

\subsubsection{High-confidence U-CW-$L_2$ Attack} We also studied the untargeted version of the high-confidence CW attack, i.e., U-CW-$L_2$ attack. Again, we use the high-confidence attack to generate adversarial examples based on the point-based classifier, since our region-based classifier is non-differentiable. For a given confidence parameter $k$, for each testing benign example, we have used the T-CW-$L_2$ attack to generate 9 adversarial examples for each candidate target label in our experiments of evaluating the high-confidence T-CW-$L_2$ attack; we use the successful adversarial example that has the smallest noise as the adversarial example generated by the high-confidence U-CW-$L_2$ attack; failure is returned if none of the 9 adversarial examples is successful. Figure~\ref{unhc} shows the success rates and average noise of successful adversarial examples for the high-confidence U-CW-$L_2$ attack, as we increase the confidence parameter $k$. For our region-based classifier, the high-confidence U-CW-$L_2$ attack has a higher success rate with a larger $k$, but the noise is also larger, except for small $k$ on the MNIST dataset. Compared with the high-confidence T-CW-$L_2$ attack, U-CW-$L_2$ attack needs a smaller $k$ to reach 100\% success rate. This is because the untargeted attack selects the best adversarial example among the adversarial examples with different target labels.

\subsubsection{Combined Evasion Attacks} Table~\ref{ensemble} shows the results for targeted and untargeted combined attacks. We described these combined attacks in Section~\ref{sec:attack}. Specifically, for each testing benign example and a target label, for T-CA-$L_0$, we use T-JSMA and T-CW-$L_0$ to generate adversarial examples,  use the successful adversarial example with the smallest noise as the adversarial example for T-CA-$L_0$, and failure is returned if none is successful; for T-CA-$L_\infty$, we use the successful adversarial example with the smallest noise, which were generated by T-FGSM, T-IGSM, and T-CW-$L_\infty$, as the adversarial example for T-CA-$L_0$, and failure is returned if none is successful;  
for T-CA-$L_2$, we use T-CW-$L_2$ with $k$ upto 40 to generate adversarial examples, select the successful adversarial example with the smallest noise as the adversarial example for  T-CA-$L_2$, and failure is returned if none is successful. Similarly, we can construct adversarial examples for untargeted combined attacks. 

From Table~\ref{ensemble}, we make several observations. First, our region-based classifier is more robust to point-based classifier, adversarial training, and defensive distillation. For instance, for the T-CA-$L_0$ attack, our region-based classifier has success rates of 55\% and 76\% on the MNIST and CIFAR-10 datasets, respectively, while all other compared classifiers have 100\% success rates. Moreover, our region-based classifier enforces an attacker to add  more noise into adversarial examples.  For instance, for the T-CA-$L_0$ attack on CIFAR-10, our region-based classifier requires 2.5 to 3 times more noise than the compared classifiers. Second, our region-based defense is less effective for untargeted combined attacks. In particular, untargeted combined attacks still achieve high success rates for our region-based classifiers. However, our region-based classifiers still enforce attackers to add larger noise to construct successful adversarial examples.  For instance, for U-CA-$L_0$ on CIFAR-10, the required noise to attack our region-based classifier is around twice of the noise required to attack the compared classifiers.

\begin{figure*}[!t]
\centering
\subfloat[MNIST]{\includegraphics[width=0.45 \textwidth]{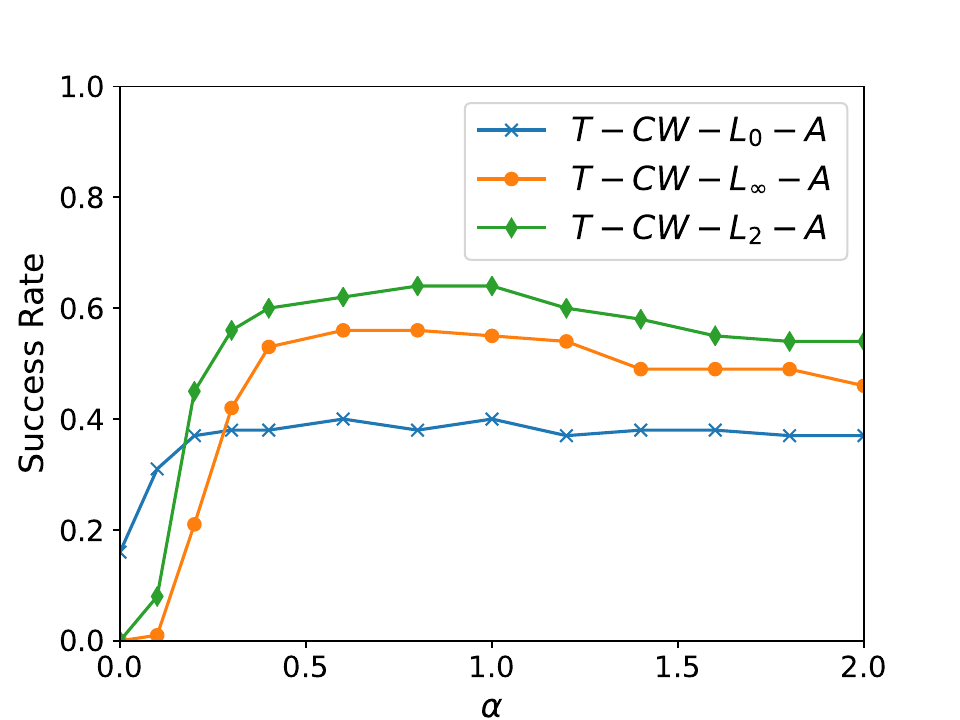}\label{mnist}}
\subfloat[CIFAR-10]{\includegraphics[width=0.45 \textwidth]{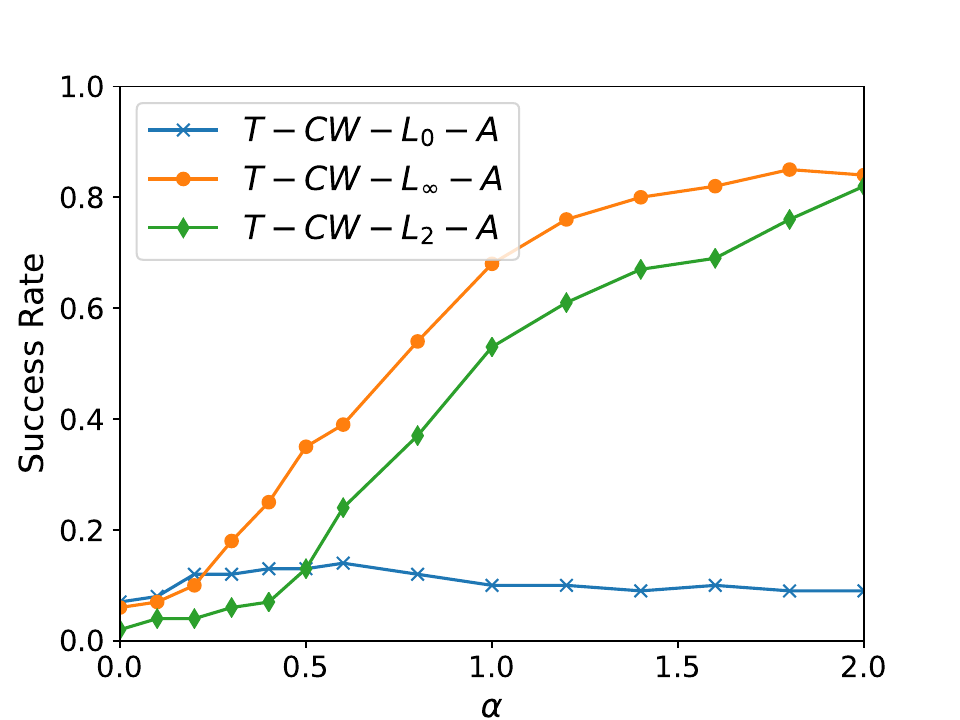}\label{cifar}}
\caption{Tradeoff between success rates and increased fraction of noise for adapted CW attacks.}
\label{adapted}
\end{figure*}

\begin{figure}[!t]
\centering
\includegraphics[width=0.45 \textwidth]{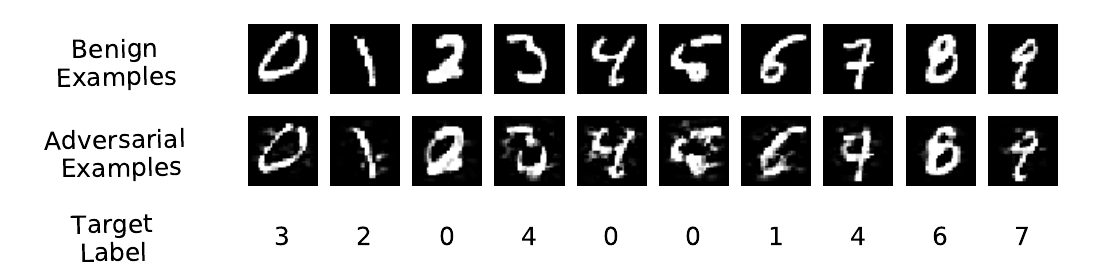}
\caption{Adversarial examples generated by the adapted targeted evasion attack T-CW-$L_2$-A for the MNIST dataset, where $\alpha=1$.}
\label{adaptedexample}
\end{figure}

\subsection{Robustness to New Evasion Attacks} Recall that we discussed adapting existing attacks to our region-based classifier in Section~\ref{sec:attack}. The key idea is to move the adversarial example further away from the classification boundary. The parameter $\alpha$ controls the tradeoff between the increased fraction of  noise (for $L_2$ and $L_\infty$ norms) and success rates. 
We focus on adapting CW attacks since they are state-of-the-art in terms of the added noise. 
Figure~\ref{adapted} shows the success rates of the adapted targeted CW attacks as we increase $\alpha$. 

The adapted targeted CW attacks cannot achieve 100\% success rates anymore no matter how we set the parameter $\alpha$. Specifically, the success rates first increase and then decrease as $\alpha$ increases. This is because adding too much noise to adversarial examples moves them to other class regions, resulting in an unsuccessful targeted evasion attack.  Suppose an adversarial example has a target label $i$. The original adversarial example generated by a targeted CW attack is in the class region $R_i$. When $\alpha$ is small, the adapted adversarial example generated by an adapted CW attack is still within the class region $R_i$. However, when $\alpha$ is large, the adapted adversarial example is moved to be in another class region $R_j$, which has a different label.

For the MNIST dataset, the best adapted targeted evasion attack is the adapted attack T-CW-$L_2$-A. The largest success rate the attack can achieve is 64\%, when $\alpha=1$, i.e., the added average noise of adversarial examples is doubled.  When the attack T-CW-$L_2$-A wants to achieve 50\% success rate, the adversarial examples need 25\% more noise. For the CIFAR-10 dataset, the best adapted targeted evasion attack is T-CW-$L_\infty$-A. The attack achieves 85\% success rates when $\alpha=1.8$. The attack needs to set $\alpha=0.75$ in order to achieve a 50\% success rate.

Figure~\ref{adaptedexample} shows the adversarial examples generated by the adapted targeted evasion attack T-CW-$L_2$-A for the MNIST dataset when the attack achieves the highest success rate, i.e., $\alpha=1$. For all these adversarial examples,    our region-based DNN classifier predicts the target label for each of them. Recall that Figure~\ref{example} shows adversarial examples generated by the existing T-CW-$L_2$ attack. To compare the adversarial examples generated by T-CW-$L_2$ and T-CW-$L_2$-A, we use the same benign examples in Figure~\ref{adaptedexample} and Figure~\ref{example}.  We observe that some adversarial examples generated by T-CW-$L_2$-A have changed the true labels. For instance, the sixth adversarial example in Figure~\ref{adaptedexample} was generated from a benign example with true label 5. However, human can hardly classify the adversarial example to be a digit 5, i.e., the true label has been changed. Similarly, the third, eighth, and ninth adversarial examples almost change the true labels of the corresponding benign examples. 

Recall that in Section~\ref{evaluationmetric}, we discussed that a successful adversarial example should satisfy two conditions and we approximate success rate of an attack using its generated adversarial examples that satisfy the second condition only. Our results in Figure~\ref{adaptedexample} show that some adversarial examples that satisfy the second condition do not satisfy the first condition, because of adding too much noises. Therefore, the real success rates of the adapted evasion attacks are even lower. 


\section{Discussions}
\label{sec:discussion}

\myparatight{Other types of regions} Our work demonstrates that, via ensembling information in a region around a testing example (benign or adversarial), we can enhance DNNs' robustness against evasion attacks without sacrificing their generalization performance on benign examples. In this work, we use a hypercube as the region. It would be an interesting work to explore other types of regions, e.g., $l_p$-norm ball (i.e., $B_p(x,r)=\{y|y_j\in [0,1] \text{ and } ||y-x||_p \leq r\}$), hypersphere, and intersection between a $l_p$-norm ball and manifolds~\cite{Narayanan:2010} formed by the natural examples. Hypercube is essentially a $l_\infty$-norm ball. Moreover, we essentially use majority vote to ensemble information in a region. It would be interesting to explore other methods to ensemble information in a region, e.g., considering certain weights for different data points in a region.

\myparatight{Randomization based defenses} Another way to interpret our region-based classification is that our method uses randomization to defend against evasion attacks. Specifically, for each testing example, our region-based classification is equivalent to add small random noise to the testing example to construct some noisy examples; we use a point-based DNN to classify each noisy example; and we take a majority vote among the noisy examples to predict the label of the testing example. We note that randomization-based defense was used as a feature preprocessing step~\cite{wang2017adversary,randomization17} to enhance robustness of DNNs. However, our work is different these defenses in two aspects. First, the randomization-based feature preprocessing needs to be used in both training and testing, while our method is only applied at testing time. Therefore, randomization-based feature preprocessing is not applicable to legacy classifiers, while our method is. Second,  randomization-based feature preprocessing applies randomization once, i.e., they essentially randomly sample one noisy example in a region around an example and use it to replace the example. Our method samples multiple noisy examples and ensembles them. 

\myparatight{Generating robust adversarial examples} Our work demonstrates that adversarial examples generated by state-of-the-art evasion attacks are not robust, i.e., if we add a small noise to an adversarial example, a classifier will very likely predict a different label for the noisy adversarial example. However, benign examples are robust to such random noise, as our region-based classifier does not sacrifice the classification accuracy. It is an interesting future work to generate adversarial examples that are robust to random noise. 

\section{Conclusion}
In this work, we propose a region-based classification to mitigate evasion attacks to deep neural networks. 
First, we perform a measurement study about adversarial examples. We observe that adversarial examples are close to the classification boundary and the hypercube around an adversarial example significantly intersects with the class region that has the true label of the adversarial example.  Second, based on our measurement study, we propose a region-based DNN classifier, which ensembles information in the hypercube around an example to predict its label. 
Third, we perform evaluations on the standard MNIST and CIFAR-10 datasets. Our results demonstrate that our region-based DNN classifier is significantly more robust to various evasion attacks than existing methods, without sacrificing classification accuracy on benign examples.    

Future work includes exploring different types of regions, different ways to ensemble information in a region, and new attacks to generate robust adversarial examples. We encourage researchers who propose new evasion attacks to evaluate their attacks against our region-based classifier, instead of standard point-based classifier only.


\bibliographystyle{ACM-Reference-Format}
\bibliography{refs,refs2}

\end{document}